\documentclass[epj,final]{svjour}
\usepackage{graphicx}
\begin{document}
\title{Quantum thermal transport in nanostructures}
\author{Jian-Sheng Wang\inst{1,}\thanks{also at Institute of High 
Performance Computing, 1 Science Park Road, Singapore 117528, and
Singapore-MIT Alliance, 4 Engineering Drive 3, Singapore 117576.}
\and Jian Wang\inst{2} \and J. T. L\"u\inst{1}}
%
%
\institute{Center for Computational Science and Engineering and Department of Physics, National University of Singapore, 
Singapore 117542, Republic of Singapore 
\and College of Physical Science and Technology, Yangzhou University, Yangzhou 225002, P. R. China}

\date{Received: 20 February 2008 / Revised version: ??} 
%
\abstract{In this colloquia review we discuss methods for thermal transport
calculations for nanojunctions connected to two semi-infinite leads served as
heat-baths.  Our emphases are on fundamental quantum theory and atomistic
models.  We begin with an introduction of the Landauer formula for ballistic
thermal transport and give its derivation from scattering wave point of
view. Several methods (scattering boundary condition, mode-matching, Piccard
and Caroli formulas) of calculating the phonon transmission coefficients are
given.  The nonequilibrium Green's function (NEGF) method is reviewed and the
Caroli formula is derived.  We also give iterative methods and an algorithm
based on a generalized eigenvalue problem for the calculation of surface
Green's functions, which are starting point for an NEGF calculation.  A
systematic exposition for the NEGF method is presented, starting with the
fundamental definitions of the Green's functions, and ending with equations of
motion for the contour ordered Green's functions and Feynman diagrammatic
expansion.  In the later part, we discuss the treatments of nonlinear effects
in heat conduction, including a phenomenological expression for the
transmission, NEGF for phonon-phonon interactions, molecular dynamics
(generalized Langevin) with quantum heat-baths, and electron-phonon
interactions.  Some new results are also shown.  We also briefly review the
experimental status of the thermal transport measurements in nanostructures.
%
} 
\maketitle
\section{Introduction}
\label{intro}
\label{sec:1}
The first quantitative description of the phenomenon of heat conduction was
given by Fourier in the early 1800s, which states that the heat current is
proportional to the temperature gradient, $J = - \kappa \nabla T$.  The
coefficient, $\kappa$, is known as the thermal conductivity.  Together with
energy conservation, the temperature field of a macroscopic body satisfies a
diffusion equation.  Debye proposed simple kinetic theory to express the
thermal conductivity in terms of a product of specific heat, velocity, and
mean free path of the phonons.  A more complete theory is given by Peierls
\cite{peierls55}, based on a Boltzmann equation for the phonons.  The
Boltzmann-Peierls equation approach for understanding the thermal transport
has been a standard approach, and has been perfected by many people
\cite{carruthers61}.  There is also recent work for a more rigorous derivation
of the Boltzmann equations \cite{spohn06}.

In recent years, much research has been done on mesoscopic to microscopic
systems, including thermal transport
\cite{cahill03,gchen04,galperin-review07}, for the obvious reason of relevance
to the miniaturization in the electronic industry.  At small scale, some of
the concepts familiar at macroscopic regime may not be applicable, e.g., the
concept of a distribution function of both coordinate and momentum used in the
Boltzmann equation.  At the nanoscale, where atomic detail becomes important,
we expect to replace the Boltzmann equation by something more fundamental.
One possibility is quantum Boltzmann equations \cite{meier69,mahan87}.
Another class of methods for thermal transport problems starting with atomic
models is molecular dynamics (MD) \cite{lepri-review03,mcgaughey06,heino07}.
The thermal conductivity can be computed in equilibrium based on the
Green-Kubo formula, or through a direct computation of the thermal current in
a nonequilibrium setting.  A variety of systems has been simulated with MD for
nanostructures and interfaces.  However, since MD is purely a classical
method, quantum effect can not be taken into account.

In this review, we consider a bottom-up approach where we start with a lattice
model with only the harmonic interactions.  Such models explain well the
thermal transport properties when systems under consideration are smaller than
their mean free path of the phonons.  This regime is known as ballistic heat
transport regime.  This will be the contents of Sec.~\ref{sec-2}. In this
section, we discuss the concept of universal thermal conductance for perfect
ballistic systems, and address the issue of computing the thermal current if
the system is not a perfect periodic lattice.  The result is given by the
Landauer formula.  Effective computational methods will be reviewed for the
transmission coefficient used in the Landauer formula.  What will happen if we
introduce interactions for the phonons?  We discuss this issue in
Sec.~\ref{sec-3}.  The problem of nonlinear interaction can be solved, at
least in principle, with the nonequilibrium Green's function (NEGF) method,
which is more fundamental than the Peierls approach, and without conceptual
difficulty.  In practice, the NEGF method is computationally extremely
intensive if the nonlinear problem is treated by mean field approximations.
We also review a new molecular dynamics method where quantum effect can be
taken into account approximately.  In particular, the ballistic regime can be
simulated correctly. Electron-phonon interaction, as well as recent
experimental progress, will be reviewed briefly.

\subsection{Models in this review\label{section-model}}

For notational consistency, in this review, we'll discuss a model of junction
connected to two leads using the following Hamiltonian
\cite{jswprb06,jswpre07}.  The model is in fact very general.  In order to
have a form for the Green's function similar to that of electrons, we use a
transformation for the coordinates, $u_j = \sqrt{m_j}\, x_j$, where $x_j$ is
the relative displacement of $j$-th degree of freedom having the dimension of
length, and we call $u_j$ the mass-normalized displacement. In this way, the
kinetic energy is always of the form $\frac{1}{2}\dot{u}^T \dot{u}$ (where the
superscript $\scriptstyle T$ stands for matrix transpose), and the mass
information is transferred completely to the spring constants.  We introduce a
superscript $\alpha$ to indicate the region.  Then $u_j^\alpha$ is the
deviation from equilibrium position for the $j$-th degree of freedom in the
region $\alpha$; $\alpha = L, C, R$, for the left, center, and right regions,
respectively. The quantum or classical Hamiltonian is given by
\begin{equation}
\label{eq-H1}
{\cal H} = \!\!\!\!\!\sum_{\alpha=L,C,R}\!\!\!\!\!H_\alpha  + (u^L)^T V^{LC} u^C + (u^C)^TV^{CR} u^R + V_n,
\end{equation}
where $H_{\alpha} = \frac{1}{2} {(\dot{u}^\alpha)}^T \dot{u}^\alpha +
\frac{1}{2} {(u^\alpha)}^T K^\alpha u^\alpha$ represents coupled harmonic
oscillators, $u^\alpha$ is a column vector consisting of all the displacement
variables in region $\alpha$, and $\dot{u}^\alpha$ is the corresponding
conjugate momentum.  $K^\alpha$ is the spring constant matrix and
$V^{LC}=(V^{CL})^T$ is the coupling matrix of the left lead to the central
region; similarly for $V^{CR}$.  The dynamic matrix of the full linear system
is
\begin{equation}
\label{eq-K}
K = \left( \begin{array}{lll} K^L & V^{LC} & 0 \\
             V^{CL} & K^C & V^{CR} \\
             0   & V^{RC} & K^R
           \end{array}\right).
\end{equation}
There will be no interaction between the two leads.  The nonlinear part of the
interaction will take the form
\begin{equation}
V_n = \frac{1}{3} \sum_{ijk} T_{ijk}\, u_i^C u_j^C u_k^C + 
\frac{1}{4} \sum_{ijkl} T_{ijkl}\, u_i^C u_j^C u_k^C u_l^C
\end{equation}
for perturbation expansion, but it can be arbitrary.  Notice that the notation
is valid independent of the actual physical dimensions used.

For easy treatment, we consider the leads as quasi-one-dimensional lattices
characterized by three square matrices of equal size, $k_{00}$, $k_{11}$, and
$k_{01}$.  The matrix $k_{00}$ is the block immediately adjacent to the
center, while $k_{11}$, and $k_{01} = k_{10}^{T}$ are repeated for the
semi-infinite chain of the lead.  They form the whole dynamic matrix of the
lead in the following way:
\begin{equation}
\label{eq-KR}
K^R = \left(\begin{array}{cccc} k_{00} & k_{01} & 0  & \cdots \\
                  k_{10} & k_{11} & k_{01} & 0 \\
                  0      & k_{10} & k_{11} & k_{01} \\
                  \cdots & 0      & k_{10} & \ddots 
           \end{array}
            \right).
\end{equation}
If there are some irregularities in the junction structure, we can enlarge the
central part, so that the lead is always a regular lattice.  The nearest
neighbor nature of $K^R$ is not a real limitation, since one can use a large
supercell so that only the neighbor cells have interactions. If the cross
sections of the leads are large, we can also model bulk walls.

The semi-infinite nature of the leads is important conceptually. First, the
heat bath by definition must be sufficiently large so that any finite energy
transfer does not affect its temperature.  Second, phonon waves scattered back
into the bath will get dissipated and will not be reflected back to the
central region.  Any finite harmonic system will not have such a property.

\section{\label{sec-2}Ballistic thermal transport}
\subsection{Landauer formula}
As early as 1957, Rolf Landauer presented a very intuitive interpretation of
electron conduction in nanoscale junctions based on the concept of wave
scattering \cite{landauer57,landauer70}.  The same argument can be applied to
the phonon transport as well
\cite{angelescu98,rego98,rego01,blencowe99,blencowe04}.  Heat current flowing
from left to right through a junction connected to two leads at different
equilibrium heat-bath temperatures $T_L$ and $T_R$ is given by the Landauer
formula
\begin{equation}
\label{eq-landauer}
I = \int_{0}^\infty\! { d \omega \over 2 \pi} \,\hbar\omega\, T[\omega] \bigl( 
f_L - f_R\bigr),
\end{equation}
where $f_{L,R} = \bigl\{\exp[\hbar \omega/(k_B T_{L,R})] - 1\bigr\}^{-1}$ is
the Bose-Einstein (or Planck) distribution for phonons, and $T[\omega]$ is
known as the transmission coefficient (or transmittance). The formula
describes the situation that the central region is small in comparison with
the coherent length of the waves so that it is treated as purely elastic
scattering without energy loss.  The dissipation resides in the heat baths.
We refer to such a situation as ballistic thermal transport. Note that the
transmission coefficient is independent of the temperature.  All the
temperature dependences are in the distribution functions $f_L$ and $f_R$.

The thermal conductance is defined as the limit
\begin{equation}
\sigma = \lim_{T_L \to T, T_R \to T} { I \over T_L - T_R}.
\end{equation}
The connection with the usual definition of thermal conductivity in the
Fourier's law, $J = - \kappa \nabla T$, is, $\kappa = \sigma L/S$, where $L$
is the length of a system in the direction of heat flow, while $S$ is the
cross-section area of the system.  For quasi-one-dimensional systems like
carbon nanotubes, $\sigma$ is a better quantity to use, since the
cross-section area $S$ is not very well defined.

\begin{figure}
\includegraphics[width=\columnwidth]{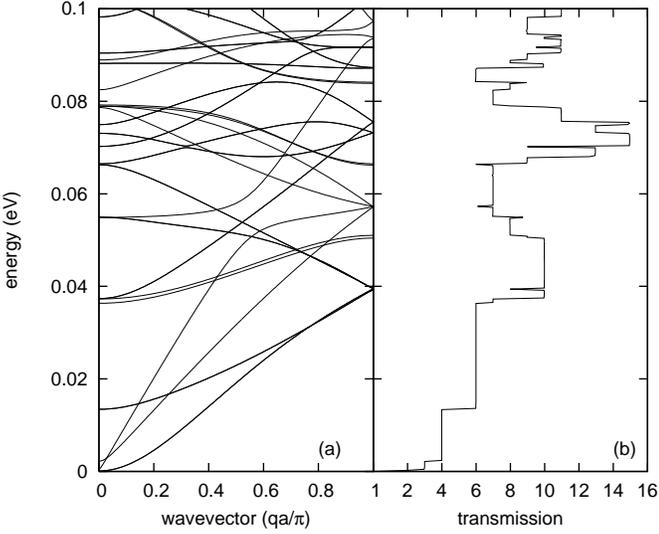}
\caption{(a) Dispersion relation $\hbar \omega$ vs.~normalized wave number of
a (6,0) carbon nanotube. (b) The phonon energy $\hbar \omega$ vs. the
transmission coefficient $T[\omega]$.}
\label{fig-disper-ballist}
\end{figure}
One very good use of the Landauer formula is that it provides an upper bound
for the thermal conductance of quasi-one-dimensional periodic systems
\cite{mingo05}.  This is obtained when all modes of waves are transmitted
without scattering or reflecting back to the heat bath.  In this ideal
situation, the transmission coefficient consists of a set of steps as a
function of $\omega$ taking only integer values, that is
\begin{equation}
\label{eq-num-mode}
T[\omega] = \mbox{{\rm number of modes at frequency $\omega$}}.
\end{equation}
Figure~\ref{fig-disper-ballist}(a) shows the dispersion relation of a carbon
nanotube of chirality (6,0) with the force constants computed from Gaussian~03
\cite{gaussian03}. We optimize a 7-period, hydrogen terminated structure. The
basis set is 6-31G, and the theoretical model is b3lyp DFT method.  Once the
dispersion relation is computed, the transmission coefficient can be obtained
by counting the numbers of positive wavevector branches of phonon vibrations.
Actually, the transmission curve in Fig.~\ref{fig-disper-ballist}(b) is
calculated by a more general NEGF method, which will be discussed in
Sec.~\ref{sec-NEGF}.

In the low-temperature limit, the high energy optical modes will not be
occupied due to the Bose-Einstein distribution.  Only the lowest energy, long
wave modes are important.  In this low-temperature limit, we have $T[\omega]
\approx N_m = 4$ for any quasi-one-dimensional systems possessing
translational invariance in three directions and rotational invariance along
the axial direction (the four acoustic modes correspond to one longitudinal
mode, two transverse modes, and an extra `twist mode'),
and the thermal conductance is given by
\begin{equation}
\sigma = \int_{0}^\infty\! { d \omega \over 2 \pi} \,\hbar\omega\, N_m 
\frac{\partial f}{\partial T} = 
N_m \frac{\pi^2 k_B^2 T}{3 h},
\end{equation}
independent of the material parameters.  This is known as the universal
quantum thermal conductance \cite{rego98}, in analogous to the universal
electron conductance $2e^2/h$.  The material constant independence stems from
the fact that in one dimension, the product of phonon density of states and
the group velocity is one.  In higher dimensions, we must integrate in the
$k$-space, instead of just a single $\omega$ \cite{wangj05}. The group
velocity enters explicitly.  The universality appears to have a deep
connection to the maximum information or entropy flow in a channel
\cite{pendry83,blencowe00}.  The approximation $T[\omega] \approx 4$ is
accurate when $k_B T$ is much smaller than the lowest optical mode energy
$\hbar \Delta \omega$.  For typical carbon nanotubes, this is valid when
temperature is of order of few kelvin to 10 K.  For Si nanowires with
transverse dimension of few hundreds nanometers, it is below 1 K.  The spacing
$\Delta \omega$ can be estimated as $\Delta \omega \approx c/\sqrt{S}$ where
$c$ is sound velocity, and $\sqrt{S}$ is an estimate to the transverse
dimension.  Indeed, Schwab {\sl et al.} have done a remarkable experiment that
verified the above theoretical prediction \cite{schwab00}.  This thermal
conductance quantum is independent of the types of energy carriers; it is also
the same for electrons \cite{chiatti06}, even photons \cite{meschke06}.

\subsection{Calculation of transmission coefficients}

When the transmission is not ideal, we must do a hard calculation for the
transmission coefficient $T[\omega]$.  Several approaches are possible.  Such
considerations also justify the Landauer formula itself.  Blencowe in
ref.~\cite{blencowe99} presented a derivation using elastic wave models.  In
this article, our emphasis will be on atomistic models.  We will give an
outline of a derivation using lattice models.  Alternatively, the Landauer
formula can be derived also from an NEGF formulism.  But before doing this, we
comment on the elastic models.

Elastic wave continuum models have been used extensively in connection with
the universal thermal conductance
\cite{angelescu98,rego98,rego01,blencowe99,blencowe04}.  This is justified on
the ground that at very low temperatures, only the long-wave, low frequency
acoustic phonons contribute to the thermal transport.  Since the wavelengths
are much larger than the lattice spacings of the materials, continuum models
are good approximations.  Often times, one takes a very simple scalar wave
governed by the equation
\begin{equation}
{ \partial^2 \Psi \over \partial t^2} - c^2 \nabla^2 \Psi = 0,
\end{equation}
with appropriate (von Neumann) boundary conditions.  The elastic wave theories
have been used to investigate the robustness of the universal thermal
conductance in various geometries including the effect of surface roughness
\cite{santamore01,santamore01prl}, abrupt junctions \cite{cross01,cmchang05},
bends \cite{sxqu04}, and other channel shapes
\cite{cleland01,tanaka04,tanaka05}.  Many papers appear recently on scalar
elastic-wave models with various geometries such as superlattices,
T-junctions, stub junctions, etc
\cite{wxli03,wxli04,wxli04apl,kqchen05,wqhuang05,wxli06,lmtang06,wxli06apl,yming06,xfpeng07,pyang07,jdlu07,ywang07,prasher07}.

It is important to ask the question, at what temperature do the elastic models
break down?  In an elastic model for a 2D channel, the dispersion relation
takes the form
\begin{equation}
\omega^2 = (cq)^2 + b n^2,\qquad n = 0,1,2,\cdots
\end{equation}
where $c$ is sound velocity, $q$ is wave number, and $b$ is some constant.
Clearly, the concept of Brillouin boundary is absent in elastic systems.
Also, the optical phonons are not represented correctly.  Thus the elastic
models will not be correct quantitatively when $k_B T > \hbar \Delta \omega$.
When the typical phonons have the wavelengths comparable to the lattice
spacings of a crystal, we expect that the continuum theory breaks down. This
happens when $\hbar c/a \sim k_BT$, where $a$ is lattice
constant. Alternatively, real crystals have a upper cut-off frequency,
$\omega_{{\rm max}}$, and associated Debye temperature, $T_{D} = \hbar
\omega_{{\rm max}}/k_B$.  The values of temperatures of the above estimates
for real systems appear to be in the range 100 to 1000 K.  It is then
interesting to see a good comparison between elastic models and lattice models
(e.g., carbon nanotubes) for thermal transport.

\subsubsection{\label{sec-derive-landauer}Scattering picture and Landauer formula}
In this subsection, we give a derivation of the Landauer formula from an
atomistic model of subsec.~\ref{section-model}, following the steps of
refs.~\cite{blencowe99,wangj05,pyang04,wangj06prb}. First, in order to define
the transmission amplitudes, we need to discuss the normal modes and the
scattering problem.  Let us consider only one lead as a quasi-one-dimensional
system consisting of $N$ repeating cells with periodic boundary conditions.
We'll take $N \to \infty$ eventually.  Due to periodicity, the displacements
of the normal modes satisfy Bloch theorem, so that all the cells are
oscillating in essentially the same way except a phase factor (see, e.g.,
ref.~\cite{ashcroft-mermin76}).  We write
\begin{equation}
\label{eq-eigen-0}
u_l = \epsilon \lambda^l e^{ -i\omega t},
\end{equation}
where $u_l$ is a column vector for the displacements in cell $l$ for $M$
degrees of freedom, $\epsilon$ is the polarization vector (of dimension $M$)
satisfying the equation
\begin{equation}
\label{eq-eigen-1}
\left(\omega^2 I - k_{11}- \frac{1}{\lambda}k_{10} - \lambda\, k_{01} \right)\epsilon = 0.
\end{equation}
The identity matrix $I$, the spring constant matrices for a single cell,
$k_{11}$, $k_{10}$, and $k_{01}$, are size $M \times M$. Imposing the Born-von
Karman boundary condition, $u_l = u_{l+N}$, one gets $\lambda = e^{i2\pi k/N}
= e^{iqa}$, $q=2\pi k/(aN)$; $a$ is the lattice constant.  We can choose the
integer $k = -N/2, -N/2 + 1, \cdots, N/2-1$ for the linearly independent
solutions.  For each $k$, Eq.~(\ref{eq-eigen-1}) gives $M$ eigenmodes with
angular frequency $\omega_{n,k}$ and polarization vector $\epsilon_{n,k}$,
where $n=1,2,\cdots$, $M$ labels the branches, and $k$ labels the wave
vectors. We'll normalize the polarization vector such that
$\epsilon_{n,k}^\dagger \epsilon_{n',k} = \delta_{n,n'}$.  The dagger stands
for Hermitian conjugate.  Note that $\epsilon_{n,k}$ in general is complex. We
demand that $\epsilon^*_{n,k} = \epsilon_{n,-k}$.  A particular vibrational
mode is considered to be $\tilde{u}_{l,n,k} = \epsilon_{n,k} e^{i2\pi kl/N}$.
A general motion can be expressed in terms of normal mode coordinates
$Q_{n,k}(t)$, as $u_l(t) = N^{-1/2} \sum_{n,k} Q_{n,k}(t) \tilde{u}_{l,n,k}$.
In terms of the normal modes, the total energy of the system is
\begin{equation}
H_R = \frac{1}{2} \sum_{n,k} \left(|\dot{Q}_{n,k}(t)|^2 + 
\omega_{n,k}^2 |Q_{n,k}(t)|^2\right).
\end{equation}

In the derivation below, we need an expression for the group velocity $v =
\partial \omega/\partial q$ in terms of the eigenvectors.  This is obtained by
differentiating the eigenvalue equation, (\ref{eq-eigen-1}), with respect to
the wavevector $q$, and multiplying the resulting equation with
$\epsilon^\dagger$ from left,
\begin{equation}
2 \omega v = i a 
\left( \lambda \epsilon^\dagger k_{01} \epsilon 
-\frac{1}{\lambda} \epsilon^\dagger k_{10}\epsilon \right).
\label{eq-group-v}
\end{equation}
We have used the Hermitian conjugate of Eq.~(\ref{eq-eigen-1}) to eliminate a
term involving $\partial \epsilon /\partial q$.

In a scattering problem setting, we usually specify a frequency first, and ask
the value of wavevector $q$ for a given mode $n$.  For this reason, we
consider the normal modes as specified by
\begin{equation}
\tilde{u}_{l,n}(\omega) = \epsilon_{n}(\omega) e^{i q_n(\omega)la}.
\end{equation}
We now introduce superscripts $(\alpha,\sigma)$ to specify the side of the
leads, $\alpha = L$ or $R$, and the direction of energy transmission $\sigma
=\pm$. We'll use the convention that energy flow into (group velocity toward)
the center is `$+$'.  The left lead and right lead can be different lattices
and thus have different dispersion relations and eigenmodes.  For notational
simplicity, we'll omit the frequency argument, and use both $(n,k)$ and
$(n,\sigma,\omega)$ to specify the modes interchangeably.  If a particular
mode $\tilde{u}_{l,n}^{(L,+)}$ is sent from the left lead to the central
region, part of it will be reflected back, and part of it will be transmitted.
For the cells far away from the center, the waves take the asymptotic form
\begin{equation}
u^L_l = \tilde{u}_{l,n}^{(L,+)} + \sum_{n'} 
\tilde{u}_{l,n'}^{(L,-)} t^{LL}_{n'n}, \quad l \to -\infty
\label{tLL-def}
\end{equation}
on the left lead, and 
\begin{equation}
u^R_l =  \sum_{n'} 
\tilde{u}_{l,n'}^{(R,-)} t^{RL}_{n'n},\quad l \to +\infty
\label{tRL-def}
\end{equation}
on the right lead.  Equation (\ref{tRL-def}) defines the transmission
amplitude $t^{RL}_{n'n}$ at frequency $\omega$ of the mode $n$ to mode $n'$ on
the right, while Eq.~(\ref{tLL-def}) defines the reflection amplitude back
into mode $n'$.  Similarly, we can send waves from the right lead to the left
lead and define analogously $t^{LR}$ and $t^{RR}$.  From these transmission
amplitudes, the total transmission, $T[\omega]$, can be calculated.

Our next task is to find a classical expression for the energy current and
then to quantize the expression and find quantum energy current.  An
expression for microscopic energy current can be obtained by considering the
conservation of energy \cite{lepri-review03}.  For a quasi-1D system, this is
\begin{equation}
{d H_l \over dt } + I_{l+1} - I_l = 0,
\end{equation}
where we define the local energy in cell $l$ as
\begin{equation}
H_l = \frac{1}{2} \left( \dot{u}_l^T \dot{u}_l 
 +  u_{l-1}^T k_{01}u_l 
 +  u_{l}^T k_{11}u_l 
 +  u_{l+1}^T k_{10}u_l \right), 
\end{equation}
such that $\sum_{l} H_l = H_{R}$ (or $H_L$).  By differentiating $H_l$ with
respective to time $t$ and using the equation of motion, we can see that
\begin{equation}
I_l = \frac{1}{2} u_l^T k_{10} \dot{u}_{l-1} 
    - \frac{1}{2} u_{l-1}^T k_{01} \dot{u}_{l} 
\label{eq-Il}
\end{equation}
satisfies the requirement. $I_l$ is the energy current from cell $l-1$ to cell
$l$.  Expressing a general vibration as superposition of modes with amplitudes
$Q_{n,k}$,
\begin{equation}
u_l(t) = \frac{1}{2\sqrt{N}} \sum_{n,k} Q_{n,k} 
\epsilon_{n,k} \lambda^l e^{-i\omega_{n,k}t} + {\rm c.c.},
\end{equation}
where c.c. stands for complex conjugate, and substituting it into
Eq.~(\ref{eq-Il}), and performing a time average, we obtain
\begin{equation}
\langle I_l \rangle = 
\frac{1}{4N} \sum_{n,k} i \omega_{n,k} |Q_{n,k}|^2 
\epsilon_{n,k}^\dagger \left(\lambda k_{01} - \frac{1}{\lambda} 
k_{10} \right) \epsilon_{n,k}.
\end{equation}
In deriving the above expression, we used the fact that the time average of
$e^{i(\omega_{n,k}-\omega_{n',k'})t}$ is zero, unless $n=n'$ and $k=k'$.
The expression in the brackets can be further simplified in terms of the group
velocity $v_{n,k}=\partial \omega_{n,k}/ \partial q$.  The final expression
for the classical energy current in terms of the normal mode amplitudes is
\begin{equation}
I = \sum_{n,k} \frac{v_{n,k}}{2aN} \omega_{n,k}^2 |Q_{n,k}|^2.
\label{eq-classical-I}
\end{equation}

We now derive a set of constraint equations for the transmission amplitudes
$t_{nn'}^{\alpha\alpha'}$ due to energy conservation.  Consider an arbitrary
linear combination of excitations at frequency $\omega$ with incident waves
from left and right, the total vibration at lead $\alpha$ can be written as
\begin{equation}
\label{eq-sum-modes}
\sum_n a_n^\alpha \tilde{u}_{l,n}^{(\alpha,+)} + 
\sum_{n,n',\alpha'} t_{nn'}^{\alpha\alpha'} a_{n'}^{\alpha'} 
\tilde{u}_{l,n}^{(\alpha,-)}.
\end{equation}
The first term above represents the superpositions of various incident modes,
while the second term contains reflected waves, as well as incident waves from
the opposite side. Comparing with the definition of normal mode amplitude, we
find $Q_{n,k} = \sqrt{N} a_n^\alpha$ for these $n$ and $k$ corresponding to
incoming waves, and equal to $\sqrt{N}\sum_{n',\alpha'}
t_{nn'}^{\alpha\alpha'} a_{n'}^{\alpha'}$ for these of outgoing waves.
Substituting the values of $Q_{n,k}$ into Eq.~(\ref{eq-classical-I}), we find
the incoming current from lead $\alpha$ to the center,
\begin{equation}
I_\alpha(\omega) = \omega^2\sum_n \frac{v_n^\alpha}{2a_\alpha} \Bigl(
|a_n^\alpha|^2 - \bigl|\sum_{n',\alpha'} t_{nn'}^{\alpha\alpha'} a^{\alpha'}_{n'}\bigr|^2 \Bigr).
\end{equation} 
In the above formula, we have the convention that $v_n^\alpha$ represents the
magnitude and thus is always positive.  Demanding that $I_L + I_R = 0$ for any
amplitudes $a_n^\alpha$, we obtain
\begin{equation}
\label{eq-t1-unitary}
{\bf t}^\dagger \tilde{\bf v} {\bf t} = \tilde{\bf v},
\end{equation}
where $\bf t$ is a matrix with elements $t_{nn'}^{\alpha\alpha'}$ where
$(\alpha,n)$ are considered row index and $(\alpha',n')$ the column index.
$\tilde{\bf v}$ is a diagonal matrix with the elements $\tilde{v}^\alpha_n =
v_n^\alpha/a_\alpha$ arranged in the same order as $\bf t$.  $a_L$ and $a_R$
are the lattice constants of the left and right lead, respectively.  The
matrix $\bf t$ is somewhat close to be unitary, but is not.  If we define
${\cal S} = \tilde{\bf v}^{1/2} {\bf t} \tilde{\bf v}^{-1/2}$, then $\cal S$
is unitary.  From ${\cal S} {\cal S}^\dagger = {\cal S}^\dagger {\cal S} = I$,
we can also show that
\begin{equation}
\label{eq-t2-unitary}
{\bf t}\tilde{\bf v}^{-1} {\bf t}^\dagger = \tilde{\bf v}^{-1}.
\end{equation}

We now discuss the quantization of the problem.  First, we consider only an
isolated lead with periodic boundary conditions. Let us introduce the
annihilation operator in Heisenberg picture $a_{n,k}(t) = a_{n,k}
e^{-i\omega_{n,k}t}$ associated with mode $(n,k)$ and its Hermitian conjugate
$a^\dagger_{n,k}(t)$ for the creation operator, satisfying the usual
commutation relations, $[a_{n,k}, a_{n',k'}] = 0$, $[a_{n,k}^\dagger,
a_{n',k'}^\dagger] = 0$, and $[a_{n,k}, a_{n',k'}^\dagger] =
\delta_{n,n'}\delta_{k,k'}$.  Then the canonical coordinate operator is
\begin{equation}
\hat Q_{n,k}(t) = \sqrt{\frac{\hbar}{2\omega_{n,k}}} \Bigl[
a_{n,k}(t) + a^\dagger_{n,-k}(t) \Bigr],
\end{equation}
satisfying $[\hat Q_{n,k}(t), \dot{\hat {Q}}{}^\dagger_{n',k'}(t)] = i \hbar
\delta_{n,n'}\delta_{k,k'}$.  Using the relation between the normal mode
coordinates and the original coordinates, we can write
\begin{equation}
\hat{u}_l(t) = \sum_{n,k} \sqrt{\frac{\hbar}{2\omega_{n,k}N}}\: 
\epsilon_{n,k}e^{i{2\pi kl\over N}} a_{n,k}(t) + {\rm h.c.},
\label{eq-u-op}
\end{equation}
where h.c. stands for Hermitian conjugate of the preceeding term.  We have
$[\hat{u}_l(t), \dot{\hat{u}}{}^T_{l'}(t)] = i\hbar I\delta_{l,l'}$, as
required. We now substitute Eq.~(\ref{eq-u-op}) into the current expression,
Eq.~(\ref{eq-Il}), taking a time average, and using the result of
Eq.~(\ref{eq-group-v}), we get
\begin{equation}
\hat{I}  = \frac{1}{aN} \sum_{n,k} v_{n,k} \hbar \omega_{n,k}
a_{n,k}^\dagger a_{n,k}.
\end{equation}
This is a well-known result for the energy current for a lattice
\cite{hardy63}.

For the open system with two leads and a central junction, we need to consider
the two sides simultaneously even if our interest is only the current on the
left lead.  Specifically the creation and annihilation operators will be
associated with the scattering state Eq.~(\ref{tLL-def}) and (\ref{tRL-def}),
originating from the left lead, $a^L_{n,k}$, and a similar one originating
from the right lead, $a^R_{n,k}$. The general displacement operator $\hat{u}$
is a superposition of the scattering waves similar to
Eq.~(\ref{eq-sum-modes}), with the number $a_n^\alpha$ replaced by the
operator $\sqrt{\hbar/(2\omega N)} a_{n,k}^\alpha(t)$ and summed over all
modes, plus its h.c.. After substituting the operator $\hat{u}$ into the
current expression, Eq.~(\ref{eq-Il}), we obtain many terms, quadratic in
$a^\alpha_{n,k}$ or ${a^\alpha_{nk}}^\dagger$.  Now at this point, great
simplification can be made if we evoke nonequilibrium thermodynamic average.
We assume that the scattering states are distributed according to equilibrium
distribution at temperature $T_L = 1/(k_B \beta_L)$ for the mode originating
from the left lead, and $T_R$ for those originating from the right lead. This
causes an asymmetry for the left moving and right moving phonon waves, thus
generates a thermal current.  The thermal average $\langle \cdots \rangle$ can
be thought of as taking the trace of the creation/annihilation operators with
respect to a density matrix given by $\rho \propto e^{-\beta_L H_L - \beta_R
H_R}$, with the following results
\begin{eqnarray}
\langle a^\alpha_{n,k} a^{\alpha'}_{n',k'} \rangle &=& 0, \\
\langle {a^\alpha_{n,k}}^\dagger {a^{\alpha'}_{n',k'}}^\dagger \rangle &=& 0,\\
\langle {a^\alpha_{n,k}}^\dagger a^{\alpha'}_{n',k'} \rangle &=& 
\delta_{n,n'}\delta_{k,k'} \delta_{\alpha,\alpha'} f_\alpha,
\end{eqnarray}
where $f_\alpha$ is the Bose-Einstein distribution function at the temperature
$T_\alpha$. Using the above results, Eq.~(\ref{eq-group-v}), and with some
algebra, we get, after cancelling the terms corresponding to the zero-point
motions,
\begin{eqnarray}
\langle I_\alpha \rangle &=& \sum_{k>0,n} 
\frac{\hbar \omega_{n,k} v_{n,k}^\alpha}{a_\alpha N_\alpha}
\Bigl[ f_\alpha(\omega_{n,k}) \qquad\qquad\qquad\qquad \nonumber  \\
&& \qquad\qquad\qquad - \sum_{n',\alpha'}|t^{\alpha\alpha'}_{nn'}|^2
f_{\alpha'}(\omega_{n,k}) \Bigr].
\end{eqnarray}
Using the properties of the matrix $\bf t$, Eq.~(\ref{eq-t1-unitary}) and
(\ref{eq-t2-unitary}), we can make the expression more symmetric with respect
to the leads.  We finally take the limit $N \to \infty$ by replacing $\sum_{k}
{v_{n,k}^\alpha}/{(a_\alpha N_\alpha)}\cdots$ by
$\int{d\omega}/{(2\pi)}\cdots$.  We obtain the Landauer formula,
Eq.~(\ref{eq-landauer}), with the total transmission given by
\cite{wangj06prb},
\begin{equation}
\label{eq-Tw-scattering}
T[\omega] = \sum_{n,n'} 
\frac{\tilde{v}^R_{n'}}{\tilde{v}^L_{n}} |t^{RL}_{n'n}|^2.
\end{equation}
The above formula has a very simple physical interpretation. The total
transmission at a given frequency is a sum of all possible incoming channels
$n$ from the left and all outgoing channels $n'$ on the right.  Each term is
simply the ratio of the energy current of the transmitted wave $\tilde
u_{l,n'}^{(R,-)}t_{n'n}^{RL}$ to the incident wave $\tilde u_{l,n}^{(L,+)}$.
Since energy must be conserved, each term cannot exceed 1.

\subsubsection{Generalized eigenvalue problem}
The eigen modes can be found by Eq.~(\ref{eq-eigen-0}) and
Eq.~(\ref{eq-eigen-1}).  In a standard dispersion relation calculation, one
specifies a wave number $q$ first, and then solves Eq.~(\ref{eq-eigen-1}) to
find various eigen frequencies.  This is inconvenient for solving a scattering
problem.  Here we like to specify a frequency $\omega$ first, and find $q$ and
associated polarization vector $\epsilon$.  To do this
\cite{khomyakov04,hzhao05,cho05}, we rewrite Eq.~(\ref{eq-eigen-1}) in a form
so that $\lambda = e^{iqa}$ becomes an eigenvalue.  Let us introduce a new
vector $\zeta$ the same dimension as $\epsilon$, by $k_{10} \epsilon = \lambda
\zeta$, then
\begin{equation}
\label{eq-eigen-2}
\left( \begin{array}{cc} \omega^2I\!\!-\!\!k_{11} & -I \\
             k_{10} & 0 
           \end{array}\right) 
\left( \begin{array}{c} \epsilon \\
             \zeta 
           \end{array}\right) = \lambda
\left( \begin{array}{cc} k_{01} & 0 \\
             0 & I 
           \end{array}\right) 
\left( \begin{array}{c} \epsilon \\
             \zeta 
           \end{array}\right),
\end{equation}
where $I$ is an $M\! \times\! M$ identity matrix.  Equation~(\ref{eq-eigen-2})
sets up a so-called generalized eigenvalue problem.  Since our eigenvalue
problem has a dimension of $2M$, we'll have upto $2M$ eigenvalues for
$\lambda$.  One can prove that the eigenvalues come in pair of inverse of each
other, $\lambda_{+} \lambda_{-} = 1$.  Most of the eigenvalues represent
evanescent modes, but these with $|\lambda| = 1$ represent traveling waves.

It is convenient to know the sign of the group velocity without actually
calculating it.  By adding a small imaginary part to $\omega$, that is,
replacing it by $\omega + i\eta$, then none of the eigenvalues $\lambda$ will
have modulus exactly 1. Considering $\eta$ as a small perturbation, we find
for the traveling waves \cite{velev04}
\begin{equation}
|\lambda| = 1 - \eta \frac{a}{v}, \quad \eta \to 0^+.
\end{equation}
That is, the forward moving waves with group velocity $v > 0$ have $|\lambda|
< 1$.  This result will be handy for constructing the surface Green's function
of the lead later.

\subsubsection{\label{sec-scattering-BC}Scattering boundary condition method}
In ref.~\cite{wangj06prb}, Wang {\sl et al.} proposed a scattering boundary
condition method to compute the transmission coefficient
$t_{n'n}^{\alpha'\alpha}$.  The method is similar but not quite the same as
the mode-matching methods \cite{ando91,khomyakov04,khomyakov05} used in
electronic transport.  The idea is very simple. We write down the linear
equations for the center, and impose the scattering boundary condition to
solve it.  Let us name the supercells consistently so that the left lead will
have $l = \cdots, -2, -1, 0$, the center consists of 1 to $N$, and right lead
is $N+1, N+2, \cdots$.  Then the equations that have not been dealt with by
the leads are
\begin{eqnarray}
\label{eq-c1}
-k_{10}^L u_{-1} + (\omega^2 - k_{00}^L) u_0 - V^{LC} u^C &=&0, \\
\label{eq-c2}
-V^{CL} u_0 + (\omega^2 - K^C) u^C - V^{CR} u_{N+1} &=& 0, \\
\label{eq-c3}
-V^{RC} u^C + (\omega^2 - k_{00}^R) u_{N+1} - k_{01}^R u_{N+2} &=&0.
\end{eqnarray}
We have assumed that the interactions between the center and leads reach only
the first layer of the leads immediately adjacent to the center.  In these
equations $u^C$ is unknown, but $u_{-1}$, $u_{0}$, $u_{N+1}$, and $u_{N+2}$
are set to the values given by Eqs.~(\ref{tLL-def}) and (\ref{tRL-def}), with
$t^{LL}_{n'n}$ and $t^{RL}_{n'n}$ treated as unknowns.  Since the form of
Eqs.~(\ref{tLL-def}) and (\ref{tRL-def}) is only asymptotic when the
evanescent modes decay to zero, we should define the center part to include
sections of the periodic leads.  Alternatively, we should also include the
evanescent modes in the summation of $n'$ in the boundary conditions, but they
should be excluded when calculating the total transmission.

However, if the evanescent modes are omitted, we have a linear system with too
many equations and few unknowns.  We solve the system numerically with a
singular-value-decomposition method \cite{NR}, which gives a very good
approximation to the original problem when the evanescent modes can be
approximated to zero.

\subsubsection{\label{sec-mode-match}Mode-matching}

Ando \cite{ando91}, Khomyakov {\sl et al.} \cite{khomyakov04,khomyakov05}, and
also Ting {\sl et al.} \cite{ting92,ting99} gave an elegant method for
computing the transmission amplitudes which we outline briefly here.  The
important idea is to eliminate $u_{-1}$ and $u_{N+2}$ in Eqs~(\ref{eq-c1}) to
(\ref{eq-c3}) so that they form a closed set of equations that can be solved
uniquely.  One first solves the generalized eigenvalue problem of
Eq. (\ref{eq-eigen-2}), then writes a general solution as linear combination
of all the modes (including the evanescent modes).  We divide the modes
according to $|\lambda|$; these modes with $|\lambda | < 1$ (forward or right
moving waves) will be designated $+$, and those of $|\lambda | > 1$ (backward
moving waves) as $-$.  We add a small positive imaginary number $i\eta$ to
$\omega$ so that none of the modes will have $|\lambda| = 1$ exactly.  Then we
can write for the perfect periodic leads
\begin{eqnarray}
\label{ul-recursion-1}
u_l &=& \sum_{\sigma = \pm} \sum_{n=1}^{M'} 
\epsilon_n^{\sigma} \lambda_{n,\sigma}^l a_n^{\sigma} \nonumber \\
& = & E^+ \Lambda^l_+ {\bf a}^+ \,+\, 
      E^- \Lambda^l_- {\bf a}^-,
\end{eqnarray}
where we have written in a more compact form by introducing matrices.  $E^+ =
(\epsilon_1^+, \epsilon_2^+, \cdots)$ is an $M \times M'$ matrix formed by the
column vectors of $\epsilon_n^+$; similarly for $E^-$. $\Lambda_{\pm}$ is an
$M' \times M'$ diagonal matrix with the eigenvalues $\lambda_{n,\pm}$, while
${\bf a}^\pm$ is a column vector for the amplitudes of the modes.  We
normalize the vectors, $(\epsilon_n^\sigma)^\dagger \epsilon_n^\sigma = 1$.
If $k_{01}$ is not singular, then $M' = M$, the number of degrees of freedom
in a cell.  However, if it is singular, then $M' < M$.  We exclude the trivial
solutions $\epsilon_{n}^\sigma=0$ in Eq.~(\ref{ul-recursion-1}).  The solution
can be split into $+$ and $-$ component, $u_l = u_l^+ + u_l^-$.  From
Eq.~(\ref{ul-recursion-1}), we see that
\begin{eqnarray}
u_{l+s}^+ &=& E^+ \Lambda^{l+s}_+ {\bf a}^+\nonumber \\
 &=& E^+ \Lambda^{s}_+ (E^+)^{-1} E^+ \Lambda^l_+ {\bf a}^+ \nonumber\\
\label{eq-ATM}
 &=& F^+(s) u_l^+,
\end{eqnarray}
where $F^+(s) = E^+ \Lambda^{s}_+ (E^+)^{-1}$.  If $M' < M$, the matrix
inverse should be interpreted as the pseudo-inverse $\bigl[(E^+)^\dagger
E^+\bigr]^{-1}\allowbreak (E^+)^\dagger$.  $F^-(s)$ is similarly defined.

With the $F$ operators defined for both the left and right lead, we can
eliminate $u_{-1}$ and $u_{N+2}$ using the equations
\begin{eqnarray}
\label{eq-u-1}
u_{-1} &=& F^{+}_L(-1)\, u_0^+ + F^{-}_L(-1)\, u_0^-, \\
\label{eq-u0}
u_0 &=& u_0^+ + u_0^-, \\
\label{eq-u+2}
u_{N+2} &=& F^{+}_R(1)\, u_{N+1}^+ = 
 F^{+}_R(1)\, u_{N+1}.
\end{eqnarray}
For the scattering state, there is no backward moving waves on the right, so
$u_{N+1}^- = 0$.  We can eliminate $u_0^-$ in favor of $u_0^+$ and $u_0$.
With that, we move the terms involving $u_0^+$ to the right-hand side of
Eq.~(\ref{eq-c1}).  For each given incident mode $u_0^+ = \epsilon_{n,L}^+$,
the transmission amplitude to $n'$ is computed by solving Eqs.~(\ref{eq-c1}),
(\ref{eq-c2}), and (\ref{eq-c3}), then obtained from the $n'$ component of a
vector
\begin{equation}
\label{eq-u+1toT}
t^{RL}_{n'n} = \left( (E_R^+)^{-1} u_{N+1} \right)_{n'}.
\end{equation}

\subsubsection{Scattering and transfer matrices}

Transfer matrix methods have been used for computing the transmission
coefficients in quasi-one-dimensional atomic models
\cite{ptong99,macia00,antonyuk04,lscao05,ymzhang05,zhangym07,murphy07} as well
as continuum models \cite{wxli03,tanaka05}. It is efficient in that the
transfer matrix can be obtained in $O(N)$ computer time where $N$ is the size
of the central region.  However, if there are evanescent modes with large
$|\lambda|$, the evaluation of the transfer matrix can be numerically rather
unstable, particularly if $N$ is large.  There is method to overcome this
problem for special cases \cite{hyin07}.

Let ${\bf a}_L^+$ be the incoming wave amplitude vector
[c.f. Eq.~(\ref{ul-recursion-1})] from the left lead, and ${\bf a}_R^-$ be the
incoming wave amplitude vector from the right lead, then the scattering matrix
relates the incoming waves to the outgoing waves
\begin{equation}
\left( \begin{array}{c}
            {\bf a}_L^- \\
            {\bf a}_R^+ 
       \end{array} \right) = 
\left( \begin{array}{cc}
  t^{LL} & t^{LR} \\
  t^{RL} & t^{RR}
       \end{array} \right) 
\left( \begin{array}{c}
            {\bf a}_L^+ \\
            {\bf a}_R^- 
       \end{array} \right), 
\end{equation}
which is the ${\bf t}$ matrix introduced in subsec.~\ref{sec-derive-landauer}.
The transfer matrix ${\cal T}$ (in the eigenmode representation), on the other
hand, relates the amplitudes of the modes of left lead to the right lead:
\begin{equation}
\label{eq-tm-T}
\left( \begin{array}{c}
            {\bf a}_R^+ \\
            {\bf a}_R^- 
       \end{array} \right) = 
\left( \begin{array}{cc}
  {\cal T}_{11} & {\cal T}_{12} \\
  {\cal T}_{21} & {\cal T}_{22} 
       \end{array} \right) 
\left( \begin{array}{c}
            {\bf a}_L^+ \\
            {\bf a}_L^- 
       \end{array} \right). 
\end{equation}
It can be checked that the transmission matrix $t^{LR} = {\cal T}_{22}^{-1}$,
or $t^{RL} = ({\cal T}^{-1})_{11}^{-1}$.  So, knowing the transfer matrix one
can deduce the transmission amplitudes.  While the scattering matrix ${\bf t}$
is always square, the transfer matrix ${\cal T}$ can be rectangular.  The
matrix inverse should be interpreted as pseudo-inverse in such a case.

We assume that the system can be divided into (principal) layers or cells,
labelled by an integer index $l$, such that only the nearest neighbors have
interactions.  Then the equations of motion at a fixed frequency $\omega$ are
given by
\begin{equation}
k_{l,l-1} u_{l-1} + \bigl(k_{l,l}-\omega^2\bigr) u_l + k_{l,l+1} u_{l+1} = 0.
\end{equation}
If the coupling matrix $k_{l,l+1}$ is not singular, we can map from the
displacements of $u_{l-1}, u_{l}$ to $u_{l}, u_{l+1}$,
\begin{eqnarray}
\!\!\!\!\!\left(\!\! \begin{array}{c}
            u_{l+1} \\
            u_{l}
       \end{array}\!\! \right) &=& 
P_{l} \left(\!\! \begin{array}{c}
            u_{l} \\
            u_{l-1}
       \end{array}\!\! \right)\nonumber \\
 &=&
\left(\! \begin{array}{cc}
   k_{l,l+1}^{-1}(\omega^2\!-\!k_{l,l})\; &\; -k_{l,l+1}^{-1}k_{l,l-1} \\
   I & 0
       \end{array}\! \right) 
\left(\!\! \begin{array}{c}
            u_{l} \\
            u_{l-1}
       \end{array}\!\! \right). 
\end{eqnarray}
We call $P_l$ also the transfer matrix (or promotion matrix).  The motions of
the left lead and right lead can be now related through
\begin{eqnarray}
\label{eq-tm-P}
\left( \begin{array}{c}
            u_{N+2} \\
            u_{N+1}
       \end{array} \right) &=&
P(N+1:0) 
\left(\! \begin{array}{c}
            u_{0} \\
            u_{-1}
       \end{array}\! \right) \nonumber \\
&=& P_{N+1} P_{N} \cdots P_1 P_0 
\left(\! \begin{array}{c}
            u_{0} \\
            u_{-1}
       \end{array}\! \right). 
\end{eqnarray}
We note that the transfer matrix for the leads, $P_l$ when $l < 0$ or $l>
N+1$, becomes independent of $l$ due to periodicity of the lattice; let us
call them $P$.  $P$ has some interesting properties: ${\rm det}(P) = 1$, and
the eigenvalue problem of $P$ is essentially the same as that of
Eq.~(\ref{eq-eigen-2}).  We can relate Eq.~(\ref{eq-tm-P}) with
(\ref{eq-tm-T}) and obtain the transmission coefficient if we relate $u_l$ to
the amplitude $\bf a$.  This is given by Eq.~(\ref{ul-recursion-1}); rewritten
in matrix form, we have
\begin{equation}
\left(\!\! \begin{array}{c}
            u_{l} \\
            u_{l-1}
   \end{array}\!\! \right) =
\left( \begin{array}{ll}
  E^+\Lambda_+^l & E^-\Lambda_-^{l} \\
  E^+\Lambda_+^{l-1} & E^-\Lambda_-^{l-1} 
       \end{array} \right) 
\left( \begin{array}{c}
            {\bf a}^+ \\
            {\bf a}^- 
       \end{array} \right). 
\end{equation}
Let us denote the transformation matrix from ${\bf a}$ to $(u_{l}, u_{l-1})^T$
by $U_L$ and $U_R$ for the left and right leads, respectively, then we have
${\cal T} = U_R^{-1} P(N+1:0) U_L$.  On the other hand, we could also use
Eq.~(\ref{eq-u-1}), (\ref{eq-u0}), and (\ref{eq-u+2}) to eliminate $u_{-1}$,
$u_{N+2}$, and solve Eq.~(\ref{eq-tm-P}) for $u_{N+1}$ and obtain the
transmission coefficient via Eq.~(\ref{eq-u+1toT}).

An interesting relation exists between the eigenvalues of `square' of the
transfer matrix and the total transmission due to Pichard \cite{pichard84}.
The results are first derived for the electronic transport where the
scattering matrix is unitary.  Since in phonon transport ${\bf t}$ is not
unitary, we cannot apply the formula directly.  Instead, we need to work with
${\cal S} = \tilde{\bf v}^{1/2} {\bf t} \tilde{\bf v}^{-1/2}$, and replace the
amplitude vector ${\bf a}$ by ${\bf c} = \tilde{\bf v}^{1/2} {\bf a}$.  Then
we have
\begin{equation}
\left( \begin{array}{c}
            {\bf c}_L^- \\
            {\bf c}_R^+ 
       \end{array} \right) = {\cal S}
\left( \begin{array}{c}
            {\bf c}_L^+ \\
            {\bf c}_R^- 
       \end{array} \right). 
\end{equation}
The corresponding transfer matrix associated with ${\cal S}$ is
\begin{equation}
{\cal M} = 
\left( \begin{array}{cc}
  {\tilde v}_R^{1/2} &  0 \\
   0         & {\tilde v}_R^{1/2} 
       \end{array} \right) 
\left( \begin{array}{cc}
  {\cal T}_{11}  & {\cal T}_{12}  \\
  {\cal T}_{21}  & {\cal T}_{22} 
       \end{array} \right) 
\left( \begin{array}{cc}
  {\tilde v}_L^{-1/2} &  0 \\
   0         & {\tilde v}_L^{-1/2} 
       \end{array} \right). 
\end{equation}
With ${\cal S}$ and ${\cal M}$ we have the Piccard identity
\cite{beenakker97,botten07}
\begin{equation}
\label{eq-piccard}
\left[ 2 I + {\cal M} {\cal M}^\dagger + ({\cal M} {\cal M}^\dagger)^{-1}
\right]^{-1} \!= \frac{1}{4}
\left( \begin{array}{cc}
  {\cal S}_{RL} {\cal S}_{RL}^\dagger & 0 \\
  0 & {\cal S}_{LR}^\dagger {\cal S}_{LR} 
       \end{array} \right), 
\end{equation}
where ${\cal S}_{\alpha\alpha'} = {\tilde v}_\alpha^{1/2} t^{\alpha\alpha'}
{\tilde v}_{\alpha'}^{-1/2}$, $\alpha = L, R$.  The eigenvalues of the product
${\cal M} {\cal M}^\dagger$ are real and positive and come in pair of inverse
of each other. We denote them by $\exp(\pm 2 x_k)$.  Taking the trace of both
sides of Eq.~(\ref{eq-piccard}), and using the fact that the total transmission
$T[\omega] = {\rm Tr} ({\cal S}_{RL} {\cal S}_{RL}^\dagger)$, we obtain
\begin{equation}
T[\omega] = \sum_{k} \frac{1}{\cosh^2 x_k}.
\end{equation}
In applying the above formula, we should use a global transfer matrix
$P(N+1+s:-s)$ for a sufficiently large $s$ to ``damp out'' the contributions
from evanescent modes.

\subsubsection{\label{sec-caroli-formula}NEGF: Caroli formula}
Another efficient way of computing the transmission coefficient is through the
so-called Caroli formula:
\begin{equation}
\label{eq-caroli}
T[\omega] = {\rm Tr}( G^r \Gamma_L G^a \Gamma_R),
\end{equation}
where $G^r = (G^a)^\dagger$ is the retarded Green's function for the central
region, while $\Gamma_{L,R}$ describes the interaction between the leads and
the central region.  Caroli {\sl et al.} first obtained a formula for the
electronic transport in a slightly more restricted case \cite{caroli71}.  Meir
and Wingreen gave the above form from NEGF formulism \cite{meir92}. For
thermal transport, it has been derived by many authors from various
perspectives
\cite{ozpineci01,segal03,mingo03,wzhang07,yamamoto06,dhar06,galperin07,jswprb06,jswpre07}.

In this subsection, we only discuss the relation of Eq.~(\ref{eq-caroli}) with
the surface Green's functions.  We'll consider a derivation in
sec.~\ref{sec-NEGF}.  The retarded Green's function of the full linear system,
including the leads, in frequency domain is given by the solution of the
equation $\bigl[(\omega + i \eta)^2I - K\bigr] G = I$.  Both $K$ and $G$ are
(infinite) matrices indexed by the labels of atomic degrees of
freedom. Partitioning the matrices according to the left, center, and right
regions, as given in Eq.~(\ref{eq-K}) for $K$, and similarly for $G$, we have
\cite{dhar06,paulsson02}
\begin{eqnarray}
\!\!\left( \begin{array}{ccc} \!(\omega\!+\!i\eta)^2I\!-\! K^L\!\! & -V^{LC} & 0 \\
             -V^{CL} & \!\!(\omega\!+\!i\eta)^2I\!-\! K^C\!\! & -V^{CR} \\
             0   & -V^{RC} & \!\!(\omega\!+\!i\eta)^2I\!-\! K^R\!
        \end{array}\right)\quad \nonumber  \\
\label{eq-partition-G}
\times \left( \begin{array}{lll} G^{LL} & G^{LC} & G^{LR} \\
                          G^{CL} & G^{CC} & G^{CR} \\
                          G^{RL} & G^{RC} & G^{RR} 
        \end{array}\right) =
\left( \begin{array}{lll} I & 0 & 0 \\
                          0 & I & 0 \\
                          0 & 0 & I
        \end{array}\right).
\end{eqnarray}
Considering only the three equations formed by each row of the first matrix
multiplying the middle column of $G$ matrix, we have
\begin{eqnarray}
\label{eq-GLC}
\bigl[ (\omega+i\eta)^2I - K^L \bigr] G^{LC} - V^{LC} G^{CC} &=& 0, \\
\!\!\!\!\!\!\!\!\!-V^{CL} G^{LC} + 
\bigl[(\omega\! +\! i \eta)^2I \!-\! K^C \bigr]G^{CC} - V^{CR}G^{RC} &=& I, \\
\label{eq-GRC}
- V^{RC} G^{CC} + \bigl[ (\omega + i\eta)^2I - K^R] G^{RC} &=& 0.
\end{eqnarray}
Introducing the retarded surface Green's functions for the leads
\begin{equation}
\label{eq-dyson-r}
g^r_\alpha = \bigl[ (\omega+i\eta)^2I -K^\alpha \bigr]^{-1}, \qquad \alpha = L, R,
\end{equation}
and using Eqs.~(\ref{eq-GLC}) and (\ref{eq-GRC}) to eliminate $G^{LC}$ and
$G^{RC}$, we can express the central part of the Green's function as
\begin{equation}
\label{eq-Gr}
G^r = G^{CC} = \bigl[ (\omega + i\eta)^2I - K^C - 
\Sigma^r_L - \Sigma^r_R \bigr]^{-1},
\end{equation}
where retarded self-energy of the leads is given by
\begin{equation}
\Sigma^r_\alpha = V^{C\alpha} g_{\alpha}^r V^{\alpha C}.
\end{equation}
$G^r$ is the matrix entering into the Caroli formula.  The $\Gamma_\alpha$
functions are given by
\begin{equation}
 \Gamma_\alpha = i (\Sigma^r_\alpha - \Sigma^a_\alpha) 
= - 2\, {\rm Im}\, V^{C\alpha} g_{\alpha}^r V^{\alpha C}.
\end{equation}
As we can see, the transmission coefficient can be obtained if we know the
surface Green's function $g^r_\alpha$. We'll discuss the algorithms for
computing the surface Green's functions in the next section.

\subsubsection{\label{subsec-surf-g}Algorithms for surface Green's functions}
In an NEGF calculation for the total transmission, an important piece of code
is for the computation of the Green's function of an isolated, semi-infinite
lead, satisfying, e.g., for the right lead,
\begin{equation}
\label{eq-g_Rr}
\bigl[ (\omega + i \eta)^2I - K^R ] g_R^r = I,
\end{equation}
where $K^R$ is given by Eq.~(\ref{eq-KR}).  In block matrix form, for the
left-most column of $g_R^r$, we have
\begin{eqnarray}
\label{eq-g00}
\bigl[(\omega + i\eta)^2 I - k_{00} \bigr]g_{00} - k_{01}g_{10} =I,\qquad\qquad \\
-k_{10} g_{l-1,0} + 
\bigl[(\omega + i\eta)^2 I - k_{11} \bigr]g_{l,0}\qquad\qquad\qquad   \nonumber \\
\label{eq-gl0}
\qquad\qquad\qquad - k_{01}g_{l+1,0} =0, 
\quad l=1, 2, \cdots
\end{eqnarray}
In order to be consistent with the literature on surface Green's functions, we
label the right lead starting from layer $l=0$ (instead of $N+1$ used in
Sec.~\ref{sec-scattering-BC}).  Since the coupling from the center to the lead
only reaches the 0-th layer, we only need the $M \times M$ block matrix
$g_{00}$.  Various methods have been used
\cite{dhlee81,guinea83,sancho84,sancho85,umerski97,sanvito99,krstic02} to find
$g_{00}$, see \cite{velev04} for an recent review.

An equation can be obtained for $g_{00}$ by considering the relation of the
original system $K^R$ and a new system with the $0$-th layer removed.  Similar
to the derivation of Eq.~(\ref{eq-Gr}), this gives
\begin{equation}
g_{00} = \bigl[ (\omega + i\eta)^2I - k_{00} - 
                              k_{01} \tilde g k_{10} \bigr]^{-1},
\end{equation}
where $\tilde g$ is the surface Green's function for a system with the 0-th
layer deleted, i.e., $\tilde g$ is the top-left element satisfying a similar
equation as Eq.~(\ref{eq-g_Rr}), but with a $K$ matrix obtained from $K^R$
with the 0-th block row and 0-th block column deleted.  If we continue this
operation of removing the next layer, we see that the $K$ matrix will not
change (since it is semi-infinite and translationally invariant), we must have
\begin{equation}
\label{eq-tilde-g}
\tilde g = \bigl[ (\omega + i\eta)^2I - k_{11} - 
                              k_{01} \tilde g k_{10} \bigr]^{-1}.
\end{equation}
The above equation can be iterated starting from $\tilde g= 0$.  This is known
as simple iteration method \cite{velev04}.  However, such iterations converge
slowly.  After $i$ iterations, the result is equivalent to a truncated system
of finite length of $i$ layers.

A more efficient and popular method is that of L\'opez Sancho {\sl et al.}
\cite{sancho85}, which doubles the number of layers in each iteration, i.e.,
after $i$ iterations, the effect of $2^i$ layers is taken into account.  We
refer to ref.~\cite{sancho85} for its derivation.  Here we give a pseudo-code
for such an algorithm.  The left arrow `$\leftarrow$' below denotes
assignment, $\epsilon$ is an error tolerance, and $| \cdot |$ stands for some
matrix norm.
\begin{verse}
$s \leftarrow k_{00}$\\
$e \leftarrow k_{11}$\\
$\alpha \leftarrow k_{01}$\\

do \{\\

\ \ \ \ $g \leftarrow [(\omega+ i\eta)^{2} I - e]^{-1}$\\
\ \ \ \ $\beta \leftarrow \alpha^T$  \\
\ \ \ \ $s \leftarrow s + \alpha g \beta$\\
\ \ \ \ $e \leftarrow e + \alpha g \beta + \beta g \alpha$\\
\ \ \ \ $\alpha \leftarrow \alpha g \alpha$ \\

\} while $(|\alpha| > \epsilon)$\\

\indent $ g_{00} \leftarrow [(\omega+ i\eta)^{2} I - s]^{-1}$.\\
\end{verse}

The most accurate and perhaps also the fastest methods
\cite{dhlee81,umerski97,sanvito99,krstic02} are based on the solution of the
generalized eigenvalue problem, Eq.~(\ref{eq-eigen-2}).  We notice that
$g_{l,0}$ and the general displacement $u_l$ satisfy exactly the same
equation, for $l > 0$.  Thus, we can express the solution of $g_{l,0}$ in the
general form of Eq.~(\ref{ul-recursion-1}).  However, since $g_{l,0}$ must be
finite in the limit $l \to +\infty$, we can only have the decay modes ($+$
components) with $|\lambda| < 1$.  In particular, Eq.~(\ref{eq-ATM}) is also
valid for $g_{l,0}$, that is,
\begin{equation}
g_{l+s,0} = F^+(s)g_{l,0},\quad F^+(s) = E^+ \Lambda^s_+ (E^+)^{-1}.
\end{equation}
Substituting this result for $l=0, s=1$ into Eq.~(\ref{eq-g00}), 
we can solve for $g_{00}$ explicitly as
\begin{equation}
g_{00} = \bigl[ (\omega + i\eta)^2 I - k_{00} - k_{01} F^{+}(1) \bigr]^{-1}.
\end{equation}
The eigenvalue problem and matrix inverses all have $O(M^3)$ in computational
complexity.  We note that when constructing the $F^+$ matrix, we need to
include all the $|\lambda| < 1$ modes, including these with $\lambda = 0$ when
$k_{01}$ is singular.
The surface Green's function of the left lead can be computed by the same
routine by first inverting the matrix indices, applying the algorithm, and
then inverting $g_{00}$ back again.

\subsubsection{Relation between scattering picture and NEGF}
The results for the transmission coefficient from the scattering picture,
Eq.~(\ref{eq-Tw-scattering}), and the Caroli formula, Eq.~(\ref{eq-caroli}),
look quite different.  It is shown explicitly in ref.~\cite{khomyakov05} that
they are completely equivalent.  This can be done by first solving formally
the mode-matching problem discussed in sec.~\ref{sec-mode-match} in terms of
Green's function, given
\begin{eqnarray}
u_{N+1} &=& G_{N+1,0} Q_0 \epsilon_{n,L}^+,\\
Q_0 &=& k_{10}^L \bigl[ F_L^+(-1) - F_L^-(-1)  \bigr].
\end{eqnarray}
The Green's function in the above formula is the top-right corner block of
$G^{RL}$ in Eq.~(\ref{eq-partition-G}), which is related to the central part
Green's function $G^r$ by
\begin{equation}
G_{N+1,0} = g^r_{R,00} V^{RC} G^r V^{CL} g^r_{L,00}.
\end{equation} 
We can relate $Q_0$ with the surface Green's functions of the leads, and
relate group velocities to self-energies, and eventually arrive at
\cite{khomyakov05}
\begin{equation}
t^{RL}_{n'n} = i2\omega \left[ \bigl(E^{+}_R \bigr)^{-1} G_{N+1,0} 
           \left((E^{-}_L)^T \right)^{-1} \right]_{n'n} \tilde{v}_n^L.
\label{eq-Gtotnn}
\end{equation}  
We can think of this result as the Fisher-Lee relation \cite{fisher81} for the
ballistic phonon transport.

\begin{figure}
\includegraphics[width=\columnwidth]{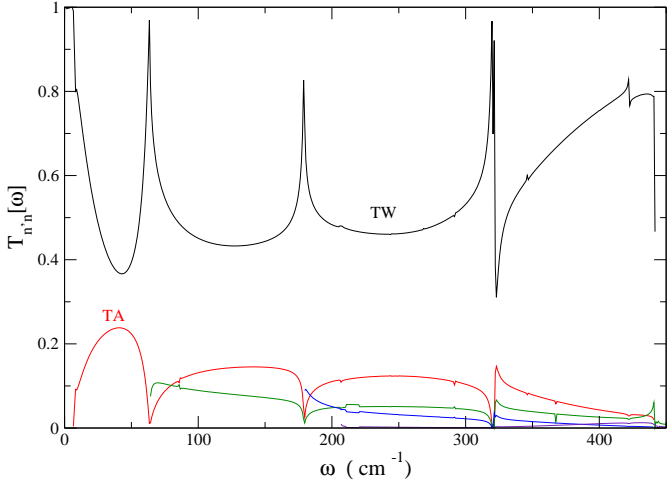}
\caption{Partial transmission $T_{n'n}[\omega]$ of the TW (twist) mode for a
(8,0) carbon nanotube with a pinning defect site.  The curve with label TW is
from TW to TW, and that with TA is from TW to TA, other unlablled curves
correspond to TW to optical modes.  The values given are sum of the two
degenerate modes of TA or optical modes.}
\label{fig-tnn}
\end{figure}

To illustrate the methods, we consider a defect model of carbon nanotube where
the mass of one of the carbon atom is set to infinite, creating a strong
pinning at the site.  The force constants are obtained from Brenner potential
\cite{brenner02}; starting with a perfect nanotube structure, the values of
rows and columns of the $K$ matrix corresponding to the defect atom are set to
0.  Unlike some of the other defects \cite{yamamoto06,kondo06,yamamoto07},
such pinning has a strong effect to the low frequency transmissions due to
broken translational invariance. It also produces strong mode mixings.  In
Fig.~\ref{fig-tnn}, we show the partial transmission, defined as,
$T_{n'n}[\omega] = |t^{RL}_{n'n}|^2\tilde{v}^R_{n'}/\tilde{v}^L_{n}$.  This
quantity can be computed by the mode-matching method of
sec.~\ref{sec-mode-match} or by Eq.~(\ref{eq-Gtotnn}).  We observe that
substantial fraction of the twist (TW) mode vibration is converted into
transverse acoustic (TA) modes and other optical modes. Due to symmetric
reason, the longitudinal mode is not excited and has zero transmission. The
peaks in the TW to TW transmission are associated with the presence of an
optical mode with zero group velocity.

\subsubsection{Transmission from molecular dynamics}

Schelling {\sl et al.} proposed a MD method to compute the transmission
coefficient \cite{schelling02,becker04,schelling04,bodapati07,kondo06}.  The
idea is to send a wave\-packet localized both in real space and momentum
space, and to measure the fraction of energy transmitted from one lead to the
other.  The method is applied to semi-conductor interfaces and nanowires,
grain boundaries, as well as carbon nanotubes with point defects.  If the
amplitudes of the waves are small enough, the results are effectively the same
as that of lattice dynamics or NEGF. Due to limited resolution in both space
and time in MD, the results are typically less accurate.  However, the
simulation can give a good visual presentation of the phonon dynamics.

The initial wavepacket can be constructed either in real space or in momentum
space. For example, we can consider the superposition of eigenmodes of
different frequencies,
\begin{equation}
u_l(t) = \int_{-\infty}^{+\infty}\!\!\!\!\!\!\! a_n(\omega) \epsilon_{n,L}^{+}(\omega)
\lambda_{n,L}^{l-l_0}(\omega) e^{-i \omega t} \frac{d\omega}{2\pi} + c.c.
\end{equation}
If we choose $a_n(\omega) \propto e^{-(\omega - \omega_0)^2/(2\sigma^2)}$,
we'll have a wave\-packet centered around $l_0$ at $t=0$ in real space with a
spread of order ${\tilde v}_n/\sigma$ lattice spacings.

\subsection{Dimension crossover, etc.}
In this section, we mention some of the work that has been done for ballistic
phonon transport, using the techniques mentioned earlier.  One interesting
aspect of ballistic transport is the dimension effect.  For
quasi-one-dimen\-sional systems, at sufficiently low temperatures, we expect
that the conductance depends on temperature linearly, $\sigma \propto T$.
Calculations on carbon nanotubes and graphene strips
\cite{yamamotoprl04,yamamoto04} confirm this expectation.  It is interesting
to know, but to our knowledge not answered, how the 1D behavior crosses over
to a full 2D behavior.  Naively, we expect that $\sigma \propto T^d$, where
$d$ is the dimensionality.  In two dimensions, this is not the case, but it is
given by $T^{3/2}$ \cite{mingo05,saito07}.  This is because that one of the
acoustic branch has a dispersion relation that is quartic in the wavevector.
In ref.~\cite{jwang07}, diameter dependence of silicon nanowires is studied
using NEGF method.  It is found that, as the diameter increases, the effective
exponent increases from 1 to 3, where $d=3$ being the bulk value. Interfaces
\cite{jwangjp07} and atomistic junction systems \cite{wzhangjht07,wzhangprb07}
have been studied.  The effect to thermal current of Stone-Wales defects in
graphene nanoribbons is studied in \cite{morooka08}; and the phonon
transmission through defects in carbon nanotubes using first principles force
constants is studied in \cite{mingo08}.  In ref.~\cite{patton01} Landauer-like
formula for thermal transport through a mesoscopic weak link is given.

\section{\label{sec-3}Nonlinear effects in thermal transport}
\subsection{Phenomenological treatment of nonlinear interactions: effective transmission}
If we take a short piece of quasi-one-dimensional material, such as a carbon
nanotube, with an ideal lead-system contact and ignoring nonlinearity, the
thermal energy transmission is given by the Landauer formula with the
transmission coefficient given by Eq.~(\ref{eq-num-mode}).  Each mode $n$ is
described by a simple rectangular function, $T_{n}[\omega] = 1$ for the
frequency $\omega$ within the phonon band of the $n$-th branch, and
$T_{n}[\omega] = 0$ outside the band.  However, this picture breaks down if
the system is long in comparison with the phonon mean free path.  In
ref.~\cite{jwangapl06} (and also \cite{murphy07}), a phenomenological
description of the effect of finite phonon mean free path is taken into
account in analogy to the electron transport \cite{dattabook1}:
\begin{equation}
\label{eq-Tn}
T_{n}[\omega] = \bigl( 1 + L/{\ell}_n(\omega) \bigr)^{-1},
\end{equation}
where $\ell_n(\omega)$ is the mean free path of phonons of mode $n$ with
frequency $\omega$ (A quite different form, $1/T[\omega] \propto (\omega -
\omega_0)^2 + d^2$, was used in ref.~\cite{qrzheng02}, which cannot 
account for the nonlinear effect).  This formula interpolates between the
ballistic transport and diffusive transport.  When $\ell_n \gg L$ we obtain
the Landauer formula, while in the opposite limit, we reproduce the well-known
Debye-Peierls formula for the thermal conductivity.

\begin{figure}
\includegraphics[width=\columnwidth]{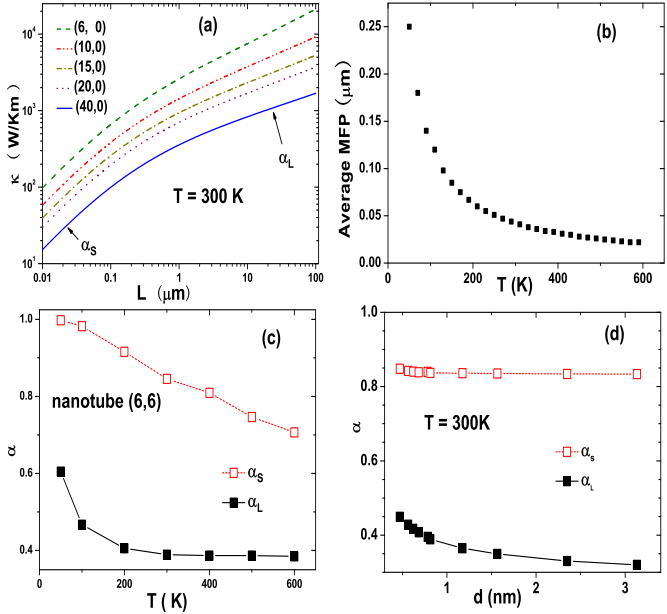}
\caption{\label{fig-jwangapl06}Carbon nanotube thermal transport in
ballistic-diffusive regimes: (a) Length dependence of thermal conductivity for
different chirality SWCNTs at 300\,K; (b) Average mean free path for nanotube
(6,6) as a function of temperature; (c) Change of the effective power-law
exponents for short and long CNTs with temperature; (d) Change of the
exponents with diameter. From \protect{ref.~\cite{jwangapl06}}.}
\end{figure}

The mean free path is related to the phonon relaxation time by $\ell_n =
\tau_n v_n$, where $v_n$ is phonon group velocity. The relaxation time can be
obtained by considering the detailed scattering mechanisms, such as Umklapp
processes, defect and boundary scatterings
\cite{callaway59,roufosse73,ecsedy77,yxiao03,jxcao04,hepplestone06,hepplestone07,yunfenggu07}.
When detailed structure and nonlinear information are put in,
Eq.~(\ref{eq-Tn}) should be able to give quantitatively correct predictions.
In ref.~\cite{jwangapl06}, a phenomenological expression $\ell \approx
A/(\omega^2 T)$ is used for a calculation of carbon nanotube. The divergence
thermal conductivity exponents are defined by $\kappa \propto L^{\alpha_S}$ or
$L^{\alpha_L}$ for short and long carbon nanotubes.  As shown in
Fig.~\ref{fig-jwangapl06}, at a fixed temperature, the thermal conductivity
grows with system sizes with a large exponent for short nanotubes and reduces
to a smaller exponent for much longer tubes.  We notice that purely diffusive
behavior is not observed even for systems as long as 100 $\mu m$.

\subsection{Nonequilibrium Green's function method\label{sec-NEGF}}

The nonequilibrium Green's function (NEGF) method is an elegant and powerful
method to treat nonequilibrium and interacting systems in a rigorous way. Its
root is from quantum field theory.  Some of the early formulations are by
Schwinger \cite{schwinger61}, Kadanoff and Baym \cite{kadanoff62}, and perhaps
most significantly by Keldysh \cite{keldysh65}.  The books of
refs.~\cite{bonitz00,bonitz03} contain some interesting historical aspects of
the developments.  The Keldysh diagrammatic expansion method is also
generalized to cases of arbitrary (correlated) initial states \cite{wagner91}.
Meir et al. \cite{meir92,jauho94} applied the technique to electronic
transport through junctions.  We recommend the books by Datta
\cite{dattabook1} and Haug and Jauho \cite{haug96} for some necessary
backgrounds in this topic.  The application of NEGF method to thermal
transport is relatively recent. In
refs.~\cite{ozpineci01,ciraci01,yamamoto06}, ballistic transport is treated.
However, the real strength of the NEGF method is its ability to handle
nonlinear interactions rigorously, at least in principle
\cite{jswprb06,jswpre07,mingo06,hpliu06,galperin07}.

\subsubsection{\label{sec-g-def}Definitions of the Green's functions and their relations}
In NEGF approach, it is convenient and also necessary to introduce several
different versions of the Green's functions.  We will start with the
definition of six Green's functions \cite{doniach74,mahan00}:
\begin{eqnarray}
G^r(t,t') &=& - i\theta(t-t') \langle [u(t), u(t')^T] \rangle, \\
G^a(t,t') &=& i\theta(t'-t) \langle [u(t), u(t')^T] \rangle, \\
G^{>}(t,t') &=& - i \langle u(t) u(t')^T \rangle, \\
G^{<}(t,t') &=& - i \langle u(t') u(t)^T \rangle^T, \\
G^{t}(t,t') &=& \theta(t-t') G^{>}(t,t') + 
              \theta(t'-t) G^{<}(t,t'),\\
G^{\bar{t}}(t,t') &=& \theta(t'-t) G^{>}(t,t') + 
              \theta(t-t') G^{<}(t,t').
\end{eqnarray}
They are known as retarded, advanced, greater, lesser, time-ordered, and
anti-time ordered Green's functions, respectively.  $u(t)$ is a column vector
of the particle displacement in Heisenberg picture. The step function
$\theta(t)=1$ if $t\ge 0$ and 0 if $t<0$.  The notation $\langle [A,
B^T]\rangle$ represents a matrix and should be interpreted as $\langle AB^T
\rangle - \langle B A^T \rangle^T$.  To avoid the $\hbar$ floating around, we
have set $\hbar = 1$ in this section.

In equilibrium or nonequilibrium steady states, the Green's functions depend
only on the difference in time, $t-t'$.  The Fourier transform of $G^r(t-t') =
G^r(t,t')$ is defined as
\begin{equation}
G^r[\omega] = \int_{-\infty}^{+\infty}\!\!\! G^r(t) e^{i\omega t}dt.
\end{equation}
Note that we use square brackets to delimit the argument for the Green's
functions in frequency domain.  The following linear relations hold in both
frequency and time domains from the basic definitions:
\begin{eqnarray}
\label{eq-1st-G-rel}
G^r - G^a &=& G^{>} - G^{<},\\
G^t + G^{\bar{t}} &=& G^{>} + G^{<},\\
G^t - G^{\bar{t}} &=& G^r + G^a.
\end{eqnarray}
Also, the relations $G^r = G^t - G^{<}$ and $G^a = G^{<}-G^{\bar{t}}$ are
useful.  Out of the six Green's functions, only three of them are linearly
independent.  However, in systems with time translational invariance, the
functions $G^r$ and $G^a$ are Hermitian conjugate of one other:
\begin{equation}
G^a[\omega] = (G^r[\omega])^\dagger.
\end{equation}
So in general nonequilibrium steady-state situations, only two of them are
independent.  We consider them to be $G^r$ and $G^{<}$, but other choices are
possible.  There are other relations in the frequency domain as well:
\begin{eqnarray}
G^{<}[\omega]^\dagger &\!=\!& - G^{<}[\omega],\\
G^r[-\omega] &\!=\!& G^r[\omega]^*,\\
\label{eq-last-G-rel}
G^{<}[-\omega] &\!=\!& G^{>}[\omega]^T \!=\! 
- G^{<}[\omega]^* \!\!+\! G^r[\omega]^T \!\!-\! G^r[\omega]^*.
\end{eqnarray}
The last two equations show that we only need to compute the positive
frequency part of the functions.
 
Equations~(\ref{eq-1st-G-rel}) to (\ref{eq-last-G-rel}) are generally valid
for nonequilibrium steady states.  In thermal equilibrium, there is an
additional equation relating $G^r$ and $G^{<}$:
\begin{equation}
G^{<}[\omega] = f(\omega) \Bigl( G^r[\omega] - G^a[\omega] \Bigr),
\label{GrtoGl}
\end{equation}
where $f(\omega)$ is the Bose-Einstein distribution function at temperature
$T$.  Equation~(\ref{GrtoGl}) is obtained by writing the Green's functions as
a sum of energy eigenstates (known as the Lehmann representation, see, e.g.,
ref.~\cite{bruus04}).  In equilibrium, we also have $G^{>}[\omega] = e^{\beta
\omega} G^{<}[\omega]$.  Thus in equilibrium, there is only one independent
Green's function; we take it to be $G^r$.

The last definition of Green's functions is the contour-ordered Green's
function, which is the most convenient entity for a diagrammatic expansion. It
is defined by
\begin{equation}
G(\tau, \tau') = -i \langle T_{\tau} u(\tau) u(\tau')^T \rangle,
\end{equation}
where the variable $\tau$ is on a Keldysh contour from $-\infty$ to $+\infty$
and back from $+\infty$ to $-\infty$.  The contour-ordered Green's function
includes four different Green's functions given earlier:
\begin{equation}
G^{\sigma\sigma'}(t,t') = \lim_{\epsilon \to 0^{+}} G(t + i\epsilon \sigma,
t' + i\epsilon \sigma'), \quad \sigma = \pm(1).
\end{equation}
We have introduced a branch index $\sigma$, such that $\tau = t + i\epsilon
\sigma$.  This is purely a book-keeping device to indicate the branch that the
$\tau$ variable is at, $\sigma = +1 $ means $\tau$ is at the $-\infty$ to
$+\infty$ branch, while $\sigma = -1$ means $\tau$ is at the returning
branch. With this notation, we can identify that $G^{++}=G^t$, $G^{--}=G^{\bar
t}$, $G^{+-} = G^{<}$, and $G^{-+} = G^{>}$, or in matrix form
\begin{equation}
\label{eq-contour-to-mat}
G(\tau, \tau') \to 
 \left( \begin{array}{ll} G^{t} & G^{<} \\
             G^{>}   & G^{\bar{t}}
        \end{array}\right).
\end{equation}

In dealing with the contour-ordered Green's functions, we often encounter
convolution of the form
\begin{equation}
B(\tau, \tau') = \!\!\int \!\!d\tau_1\!\! \int\!\! d\tau_2 \cdot\cdot 
A_1(\tau, \tau_1) A_2(\tau_1, \tau_2) \cdot \cdot
A_n(\tau_{n-1}, \tau').
\end{equation}
This form of expression can be easily translated into the retarded and lesser
Green's functions in frequency domain by the Langreth theorem as
\cite{langreth76,haug96,nzeng07}
\begin{eqnarray}
B^r[\omega] &=& A_1^r[\omega] A_2^r[\omega] \cdots A_n^r[\omega],\quad n=2,3,\cdots \\
\label{eq-lengreth-less}
B^{<}[\omega] &=& A_1^r [\omega]  \cdots A_{n-1}^r[\omega] A_n^{<}[\omega] + \nonumber \\
 && A_1^r [\omega]  \cdots A_{n-2}^r[\omega] A_{n-1}^{<}[\omega] A_n^{a}[\omega] + \nonumber \\
 && \cdots + A_1^{<} [\omega] A_2^{a}[\omega]  \cdots A_{n-1}^{a}[\omega] A_n^{a}[\omega].
\end{eqnarray}

\begin{figure}
\includegraphics[width=\columnwidth]{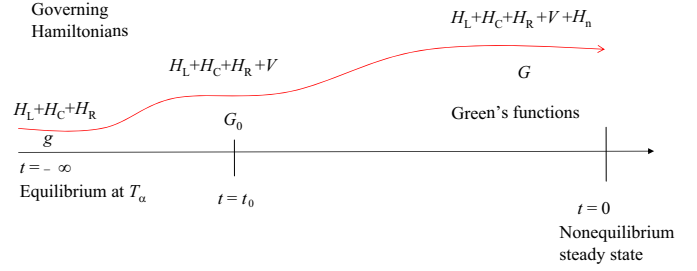}
\caption{\label{fig-adiabatic}A schematic to illustrate the two adiabatic
switch-ons. From ref.~\protect{\cite{jswpre07}}.}
\end{figure}

\subsubsection{Green's functions of the free leads}

We will adopt the following notations, the Green's functions of the
semi-infinite, equilibrium free left or right lead will be denoted $g$, while
the ballistic system of nonequilibrium, noninteracting Green's functions will
be denoted $G_0$, and the full interacting system will be denoted by $G$.
Their relations are shown schematically in Fig.~\ref{fig-adiabatic} through
two adiabatic switch-on's of the interactions.  In this subsection, we discuss
the Green's function of the leads.  To make a connection with the Green's
functions discussed in sec.~\ref{sec-caroli-formula}, we consider the
equations of motion of the Green's functions of the right lead.  By
differentiating the definitions of various Green's functions with respect to
time $t$ twice, using the Heisenberg equation of motion of any dynamic
variable $A(t)$ of the right lead,
\begin{equation}
i\frac{d A(t) }{dt} = [A(t), H_R], 
\end{equation}
we can obtain the following set of equations,
\begin{eqnarray}
{ \partial^2 g^{r,a,t}(t,t') \over \partial t^2}  + K^R g^{r,a,t}(t,t') &=& -I \delta(t-t'),\\
{ \partial^2 g^{\bar{t}}(t,t') \over \partial t^2}  + K^R g^{\bar{t}}(t,t') &=& I \delta(t-t'),\\
{ \partial^2 g^{<,>}(t,t') \over \partial t^2}  + K^R g^{<,>}(t,t') &=&0.
\end{eqnarray}
We can also give an equation for the contour ordered version,
\begin{equation}
{ \partial^2 g(\tau,\tau') \over \partial \tau^2}  + K^R g(\tau,\tau') 
= -I \delta(\tau,\tau'),
\end{equation}
where the $\tau$ variable takes values on the Keldysh contour.  The contour
ordered version is related to the ordinary Green's functions through
Eq.~(\ref{eq-contour-to-mat}).  In order to be consistent with the equations
for $g^t$ and $g^{\bar{t}}$, we must define a generalized $\delta$-function on
the contour as
\begin{equation}
\label{eq-g-delta}
\delta(\tau ,\tau') \to \sigma \delta_{\sigma,\sigma'} \delta(t-t').
\end{equation}

Solutions to these differential equations can be obtained by Fourier
transform. For example, we have
\begin{equation}
\left( \omega^2 - K^R \right) g^{r,a,t}[\omega] = I.
\end{equation}
Although $g^r$, $g^{a}$, and $g^{t}$ satisfy the same equation, their
solutions are clearly different.  They have different boundary conditions.
The correct choice for the retarded Green's function $g^r$ is
\begin{equation}
\label{Gr-free}
g^r[\omega] = \bigl[ (\omega + i \eta)^2 - K^R \bigr]^{-1},
\end{equation}
where $\eta$ is an infinitesimal positive quantity to single out the correct
path around the poles when performing an inverse Fourier transform, such that
$g^r(t) = 0$ for $t<0$.  Other Green's functions can be obtained through the
general relations among the Green's functions, e.g., $g^{<}[\omega] =
f(\omega) \bigl( g^r[\omega] - g^a[\omega] \bigr)$.

\subsubsection{Green's function $G_0$ and $G$}

To compute the Green's functions of the nonequilibrium and interacting
systems, we need to use perturbation theory and the concept of adiabatic
switch-on.  We can think of two adiabatic switch-on's as illustrated in
Fig.~\ref{fig-adiabatic}.  We imagine that at $t = -\infty$ the system has
three decoupled regions, each at separate temperatures, $T_L$, $T_C$, and
$T_R$.  The couplings between the regions and the nonlinear interactions are
turned off.  The equilibrium Green's functions $g^\alpha$ at temperature
$T_{\alpha}$ are known and take the form of Eq.~(\ref{Gr-free}).  The
couplings $V^{LC}$ and $V^{CR}$ are then turned on slowly, and a steady state
of the linear system is established at some time $t_0 \ll 0$.  For this linear
problem, the result does not depend on $T_C$; the initial condition of the
finite center part is forgotten.  Finally, the nonlinear interaction $V_n$ is
turned on, and at time $t=0$, a nonequilibrium steady state is established.

The density matrices at time $t=-\infty$, $t_0$, and $t=0$ are related in the
following way in the (respective) interaction pictures:
\begin{eqnarray}
\rho(t_0) &=&  S_0(t_0,-\infty) \rho(-\infty) S_0(-\infty, t_0),\\
S_0(t,t') &=& {\cal T} e^{-i \int_{t'}^{t} V(t'')dt''},\\
\rho(0) &=& S(0,t_0) \rho(t_0) S(t_0,0),\\
S(t,t') &=& {\cal T} e^{-i \int_{t'}^{t} V_{n}(t'')dt''},
\end{eqnarray}
where $\cal T$ is the time-order operator (assuming $t > t'$). 

The contour-ordered Green's function can be obtained by a perturbation
expansion of the interaction picture evolution operators (scattering matrix
operators) and is expressed as:
\begin{eqnarray}
\label{eq-expand-0}  
G_{0}(\tau, \tau') 
&=&  - i \langle {\cal T}_{\tau} u(\tau) u^{T}(\tau') 
e^{-i \int V(\tau'')d\tau''}  \rangle_{g},\\
\label{eq-expand-1}  
G(\tau, \tau') 
&=&  - i \langle {\cal T}_{\tau} u(\tau) u^T(\tau') 
e^{-i \int V_n(\tau'')d\tau''}  \rangle_{G_0},
\end{eqnarray}
where in Eq.~(\ref{eq-expand-0}) the unperturbed system is the uncoupled
system with Green's function $g$, while in Eq.~(\ref{eq-expand-1}) we make
perturbative expansion from $G_0$.  Since the coupling $V$ is quadratic, its
expansion leads to exact results, e.g., Dyson equation for the central part of
variables (dropping the superscript $\scriptstyle CC$),
\begin{equation}
G_0(\tau,\!\tau')\! = \! g^C(\tau,\!\tau')\! +\! 
\int\!\! d\tau_1 d\tau_2\, g^C(\tau,\!\tau_1) \Sigma(\tau_1,\!\tau_2) G_0(\tau_2,\!\tau'),
\label{dyson0}
\end{equation}
where the self-energy due to linear interactions with the leads is
\begin{equation}
\label{eq-self-energy-linear}
\Sigma(\tau_1, \tau_2) = V^{CL}g^L(\tau_1, \tau_2) V^{LC} + 
V^{CR}g^R(\tau_1, \tau_2) V^{RC}.
\end{equation}
Thus, the Green's function $G_0$ can be expressed in terms of $g$ exactly.
Using the Langreth theorem discussed at the end of Sec.~\ref{sec-g-def}, the
contour ordered Dyson equation gives two independent equations, the retarded
version has solution identical to Eq.~(\ref{eq-Gr}), while the lesser
component can be solved to given \cite{haug96}
\begin{equation}
G_0^{<}[\omega] = G_0^r[\omega] \Sigma^{<}[\omega] G_0^{a}[\omega].
\end{equation}

However, for the nonlinear problem of the second expansion, we need to use the
machinery of Feynman diagrammatic technique
\cite{abrikosov63,fetter71,zagoskin98}.  Since the nonlinear interactions 
depend on the displacement $u^C$, we only need to consider the Green's
functions $G^{CC}$.  By expanding the exponential in Eq.~(\ref{eq-expand-1}),
a series in the nonlinear interaction strength is obtained.  The Feynman
diagrams generated here are identical to the field theories of $\phi^3$ and
$\phi^4$ \cite{amit84}, and are similar to that in
refs.~\cite{procacci92,valle92,valle00} for phonon retarded self-energies.
Since the structure of the diagrammatic expansion of the contour ordered
Green's functions is identical to the standard ground state ($T=0$) expansion
for the time-ordered Green's function, the Wick's theorem \cite{fetter71} is
also applicable here.  The result of each term in the expansion can be
expressed as product of $G_0$.

\begin{figure}
\includegraphics[width=\columnwidth]{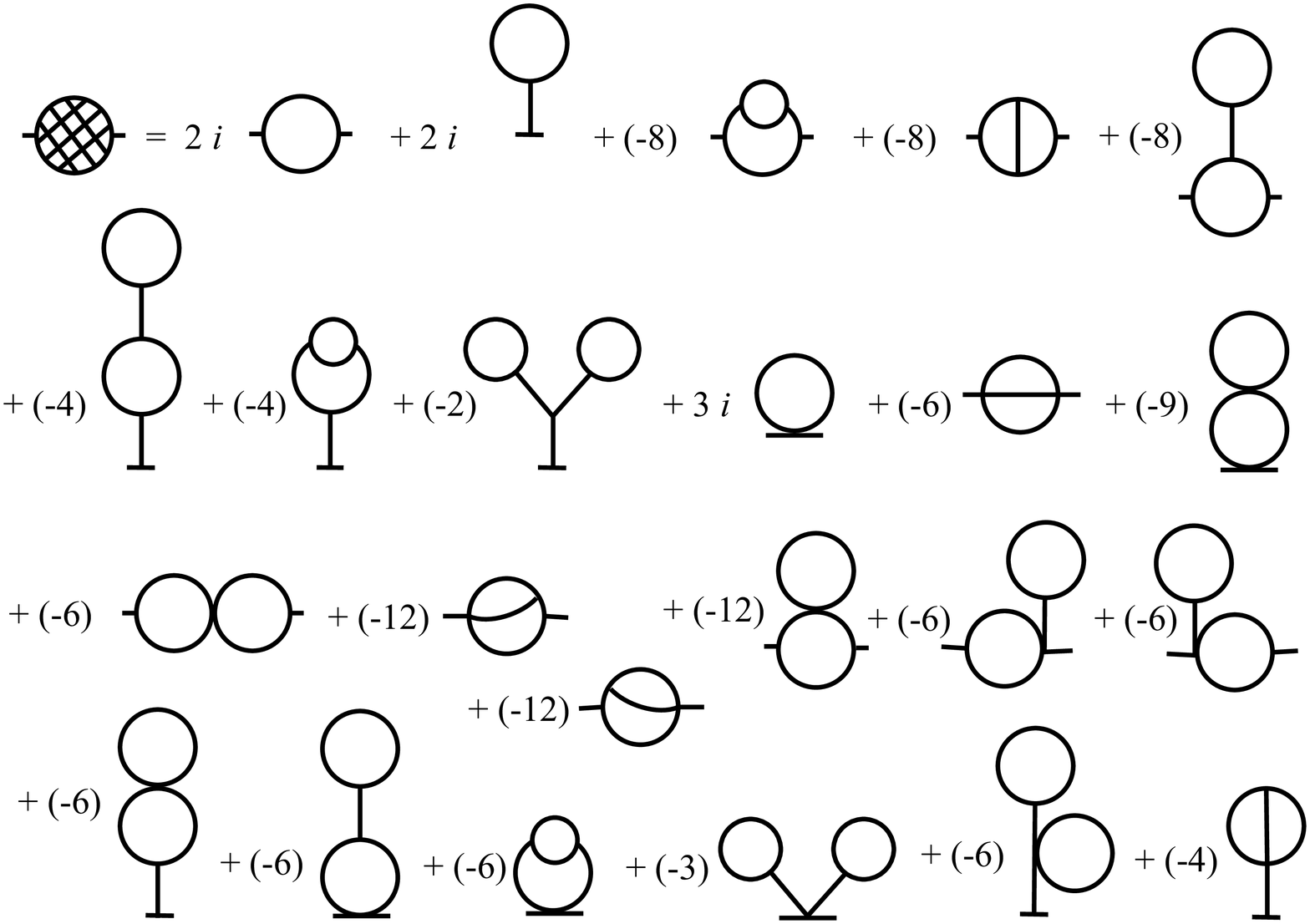}
\caption{\label{fig-feynmant4}The Feynman diagrams for nonlinear self-energy
$\Sigma_n$ with the prefactors for graphs of cubic and quartic interactions to
second order in $\hbar$.  From ref.~\protect{\cite{jswpre07}}.}
\end{figure}

The expansion contains connected as well as disconnected Feynman diagrams.
The disconnected diagrams are constant in time (not necessarily zero), and
give rise to a thermal expansion effect.  We can show that these diagrams do
not contribute to the thermal transport, as they are proportional to
$\delta(\omega)$ in the frequency domain.  The thermal current formula has a
factor of $\omega$ which makes it zero.  The connected part of the Green's
function satisfies a similar contour-ordered Dyson equation \cite{mahan00}
relating $G_c$ to $G_0$ through a nonlinear self-energy $\Sigma_n$:
\begin{equation}
G_c(\tau,\!\tau')\! = \! G_0(\tau,\!\tau')\! +\! \!
\int\!\! d\tau_1 d\tau_2 G_0(\tau,\!\tau_1) \Sigma_n(\tau_1,\!\tau_2) G_c(\tau_2,\!\tau').
\label{dyson}
\end{equation}
In ordinary Green's functions and in frequency domain ($\omega$ argument
suppressed), the above Dyson equation has solutions \cite{haug96}:
\begin{eqnarray}
G_c^r &\!=\!& \left( 
(\omega+i\eta)^2 I - K^C  -\Sigma^r 
 - \Sigma^r_n \right)^{-1}, \\
G_c^{<} &\!=\!& G_c^r \bigl(\Sigma^{<} + \Sigma^{<}_n\bigr) G_c^a .
\label{eq-G4}
\end{eqnarray}
The Eq.~(\ref{eq-G4}) is known as the Keldysh equation.  The connected part of
the Green's function $G_c$ is needed to compute the heat current.  The extra
subscript $c$ will be dropped from this point.

\subsubsection{Equation of motion method}

An alternative method to obtain the Green's functions of interacting systems
is through the equation of motion of the Green's functions.  The equations of
motion for various Green's functions may be used as a starting point to form
mean-field type approximations.  In ref.~\cite{niu99} the equation of motion
for the nonequilibrium Green's function is justified from the Keldysh
formalism, so the two approaches are completely equivalent.

It is convenient to work directly on the contour ordered Green's functions. To
represent the ordering along the Keldysh contour, we introduce a generalized
$\theta$-function \cite{leeuwen06}, $\theta(\tau, \tau')$, which is defined to
be 1 if $\tau$ is later than $\tau'$ along the contour and 0 otherwise.  Its
derivative along the contour is the generalized $\delta$-function,
\begin{equation}
\label{eq-def-delta}
\delta(\tau, \tau') = {\partial \theta(\tau, \tau') \over \partial \tau }.
\end{equation}
The generalized $\delta$-function is related to its $(\sigma, \sigma')$
components by Eq.~(\ref{eq-g-delta}).  The retarded, advanced, and
time-ordered components of $\delta(\tau, \tau')$ is the same as the ordinary
$\delta(t-t')$, lesser and greater component is 0, while anti-time-ordered
component is $-\delta(t-t')$.  We also need to generalize the canonical
commutation relation on the contour, as
\begin{equation}
\label{eq-commutator}
\Bigl[ u_j(\tau),  {d \over d \tau} u_l(\tau) \Bigr] 
= i\, \delta_{jl}.
\end{equation}
Similarly, we also generalize the Heisenberg equation of motion,
\begin{equation}
\label{eq-heisenberg}
{ d A(\tau) \over d\tau } = i [ H(\tau), A(\tau) ].
\end{equation}
Equations (\ref{eq-commutator}) and (\ref{eq-heisenberg}) are the same as the
usual canonical commutation relation and the Heisenberg equation of motion on
the upper branch ($-\infty$ to $\infty$) of the Keldysh contour; but on the
lower return branch, they are the time-reversed version of the equations
(since $d\tau = \sigma dt$).

We first consider the linear case, $V_n=0$.  If we view the system as a whole,
the contour ordered Green's function satisfies
\begin{equation}
\label{eq-G0}
-{ \partial^2 G_0(\tau,\tau') \over \partial \tau^2}  - K G_0(\tau,\tau') 
= I \delta(\tau,\tau').
\end{equation}
This is obtained from taking derivatives twice to the definition of the
contour-ordered Green's function, and using Eqs.~(\ref{eq-def-delta}) and
(\ref{eq-commutator}).  If we partition the matrix $G_0$ to the submatrices
$G^{\alpha, \alpha'}_0$, $\alpha, \alpha'= L, C, R$, and similarly for $K$, we
can obtain an equation similar in structure to Eq.~(\ref{eq-partition-G}) with
$\omega^2$ replaced by $-I {\partial^2 \over \partial \tau^2}$ and $I$ by $I
\delta(\tau, \tau')$ on the right-hand side of Eq.~(\ref{eq-partition-G}).
Using a similar solution technique, we can solve for the contour-ordered
nonequilibrium Green's functions
\begin{equation}
\label{eq-GCL}
G^{LC}_0(\tau, \tau') = \int d\tau'' g^L(\tau, \tau'') V^{LC} G^{CC}_0(\tau'', \tau'),
\end{equation}
\begin{eqnarray}
G^{CC}_0(\tau, \tau') &=& g^{C}(\tau, \tau') + \nonumber \\
&& \!\!\!\!\!\!\! \!\!\!\!\!\!\! \!\!\!\!\!\!\! \!\!\!\!\!\!\! 
\int d\tau_1 \int d\tau_2
\,g^{C}(\tau, \tau_1) \Sigma(\tau_1, \tau_2) G^{CC}_0(\tau_2, \tau').
\end{eqnarray} 
The first of the above equations relates the central part of the Green's
functions to the lead-center Green's function, which, it turns out, is also
valid when the nonlinear interaction in the center is nonzero.  The second
equation is the Dyson equation relating the free leads to coupled linear
system. The self-energy $\Sigma(\tau_1, \tau_2)$ is given by
Eq.~(\ref{eq-self-energy-linear}).

For the expansion of the Green's function $G$ in terms of $G_0$, we need to
define a general $n$-point contour-ordered Green's function as
\begin{eqnarray}
G_{j_{1}, j_{2}, \cdots, j_{n}}^{\alpha_{1}, \alpha_{2}, \cdots, \alpha_{n}}
(\tau_1, \tau_2, \cdots, \tau_n) = \qquad\qquad\qquad\qquad \nonumber \\
\qquad\qquad\qquad - i \langle {\cal T}_{\tau} u_{j_1}^{\alpha_1}(\tau_1) 
u_{j_2}^{\alpha_2}(\tau_2) \dots 
u_{j_n}^{\alpha_n}(\tau_n)\rangle.
\end{eqnarray}
The index $\alpha =L, C, R$ labels the region, $j$ labels the degrees of
freedom in that region, and $\tau$ is the contour variable.  The function is
symmetric with respect to simultaneous permutations of the triplet $(\alpha,
j, \tau)$.

We can write the interactions in a more symmetric form, e.g., for the cubic
term,
\begin{eqnarray}
\!\!\!\!\!\!\!\int V_n(\tau) d\tau &=&  \nonumber \\
 && \!\!\!\!\!\!\!\!\!\!\!\!\!\!\!\!\!\!\!\! 
\frac{1}{3} \sum_{ijk} \int\int\int T_{ijk}(\tau, \tau', \tau'') \nonumber \\
 && \qquad u_{i}^C(\tau) u_{j}^C(\tau')u_{k}^C(\tau'')\, d\tau d\tau' d\tau''.
\end{eqnarray}
Since the contour integral is $\int d\tau = \sum_{\sigma}
\sigma\int_{-\infty}^{+\infty} dt$, we must define
\begin{eqnarray}
T_{ijk}(\tau,\tau',\tau'') &=& T_{ijk} \delta(\tau, \tau') \delta(\tau, \tau'')
\nonumber \\
&\to& T_{ijk}
\sigma' \delta_{\sigma,\sigma'} \delta(t-t') 
\sigma'' \delta_{\sigma,\sigma''} \delta(t-t''). \qquad
\end{eqnarray}

\begin{figure}
\includegraphics[width=\columnwidth]{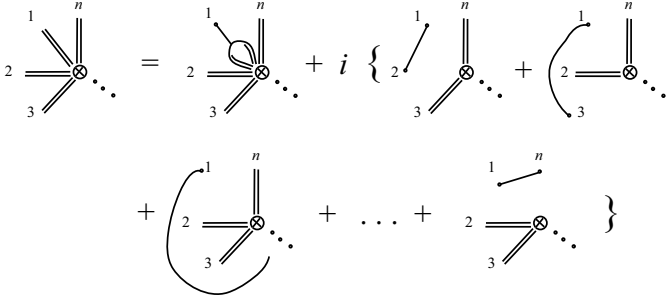}
\caption{\label{fig-exprule}Recursive expansion rule for Green's functions. A
vertex with $n$ double lines denotes an $n$-point Green's function.  The
single line denotes $G_0$. A single-line three-terminal vertex is associated
with $T(\tau,\tau',\tau'')$.  From ref.~\protect{\cite{jswpre07}}.}
\end{figure}

We now consider the equation of motion of the central part Green's function
$G^{CC}$ when there is a cubic nonlinear interaction.  Differentiating the
Green's function with respect to $\tau$ twice, using the Heisenberg equation
of motion (on the contour $\tau$), we get
\begin{eqnarray}
{\partial^2 G^{CC}_{jl}(\tau, \tau') \over \partial \tau'^2} 
+ \sum_{k}G^{CC}_{jk}(\tau, \tau') K^C_{kl} = 
- \delta_{jl} \delta(\tau , \tau') \nonumber\\
-  \sum_{k} G^{CL}_{jk}(\tau, \tau') V^{LC}_{kl} 
-  \sum_{k} G^{CR}_{jk}(\tau, \tau') V^{RC}_{kl}  \nonumber\\
- \sum_{ik} \int\!\!\int G^{CCC}_{jik}(\tau,\tau_1,\tau_2)T_{ikl}(\tau_1, \tau_2, \tau')  d\tau_1 d\tau_2.
\quad\label{eqG_CC}
\end{eqnarray}
The Green's functions $G^{CL}$ and $G^{CR}$ can be eliminated in favor of
$G^{CC}$ through (transpose of) Eq.~(\ref{eq-GCL}). Then the
differential-integral equation for $G^{CC}$ can be solved in terms of
$G^{CCC}$ and $G_0^{CC}$ with the help of Eq.~(\ref{eq-G0}).  We obtain
\begin{eqnarray}
G^{CC}_{jl}(\tau, \tau') =  G_{0,jl}^{CC}(\tau,\tau') +  
\sum_{ikm} \int\!\!\int\!\!\int
G_{0,ji}^{CC}(\tau,\!\tau_1) \qquad \nonumber \\
T_{ikm}(\tau_1,\! \tau_2,\! \tau_3)  
G^{CCC}_{kml}(\tau_2,\!\tau_3,\!\tau')
d\tau_1 d\tau_2 d\tau_3.\qquad
\label{eqG_CC_rule}
\end{eqnarray}
This equation can be represented by a diagram.  Since $G^{CC}$ is related to a
three-point Green's function, we also need the equation of motion of
$G^{CCC}$.  We can derive general rules \cite{jswpre07} for $n$-point Green's
functions, summarized as follows (as shown in Fig.~\ref{fig-exprule}):

(1) Replace leg 1 by inserting a nonlinear coupling $T(\tau, \tau', \tau'')$
such that the outer leg is immediately connected with $G_0$ while the other
two terminals increase the order of the Green's function by 1.  Quartic
interaction is similar, but the process will increase the order by 2.

(2) Add imaginary unit $i$ times a sum of the $n-1$ graphs formed by pairing
each leg with leg 1 and connecting with the propagator $G_0$, multiplied by a
$(n-2)$ order remaining Green's function.

(3) Symmetrize the graphs, if desired, i.e., do steps (1) and (2) for every
leg, 1, 2, $\cdots$, $n$, add them up, and then divide by $n$.

These rules can be programmed with a symbolic language such as {\sc
Mathematica}.  The results are identical to the usual expansion of $\exp[-
i\int V_n(\tau) d\tau]$ and applying the Wick theorem.

\subsubsection{Heat current and conductance}

The formulism discussed above offers us computational method for the Green's
functions, but the most important quantity to calculate in thermal transport
is the heat current.  In this subsection, we discuss the relationship between
the Green's functions and heat current or thermal conductance.  The
derivation of heat current formula is similar to that of electron current in
\cite{meir92,haug96}.  For interacting phonon systems, this has been carried
out in \cite{jswprb06,jswpre07,mingo06,galperin07}.

The average rate of energy decrease in the left lead is the current flow from
the left lead to the central region,
\begin{equation}
I_L = - \langle \dot{H}_L(t) \rangle.
\end{equation}
In steady state, energy conservation means that $I_L + I_R = 0$.  Using the
Heisenberg equation of motion, we obtain, at $t=0$, $I_L = \langle
(\dot{u}^L)^T V^{LC} u^C \rangle$.  The expectation value can be expressed in
terms of the Green's function $G^{<}_{CL}(t,t') = - i \langle u^L(t') u^C(t)^T
\rangle^T$. Since operators $u$ and $\dot{u}$ are related in Fourier space as
$\dot{u}[\omega] = -i\,\omega u[\omega]$, we can eliminate the derivative and
get,
\begin{equation}
I_L = - \frac{1}{2\pi} \int_{-\infty}^\infty \!\!\!\!{\rm Tr}\left(
V^{LC} G^{<}_{CL}[\omega]\right) \omega\, d\omega.
\end{equation}
Using the result derived earlier relating the mixed lead-center Green's
function to the center-only Green's function (transpose of Eq.~(\ref{eq-GCL})
with $G_0$ replaced by $G$) and applying the Langreth theorem,
Eq.~(\ref{eq-lengreth-less}), we have $G^{<}_{CL}[\omega] = G^r_{CC}[\omega]
V^{CL} g^{<}_L[\omega] +G^{<}_{CC}[\omega]V^{CL}g^{a}_L[\omega]$.  The
expression for the energy current is then given by
\begin{equation}
I_L = - \frac{1}{2\pi}\!\! \int_{-\infty}^{+\infty}\!\!\!\!\!\!\!\! d\omega\, \omega
\, {\rm Tr}\Bigl( G^r[\omega] \Sigma^{<}_L[\omega] + 
G^{<}[\omega] \Sigma^a_L[\omega] \Bigr),
\label{eq-heat-current}
\end{equation}
where $\Sigma_L(\tau,\tau') = V^{CL} g_{L}(\tau,\tau') V^{LC}$.  For
notational simplicity, we have dropped the subscript $C$ on the Green's
functions denoting the central region.  We can obtain a symmetrized expression
with respect to left and right lead and make it explicitly real,
\begin{eqnarray}
I = \frac{1}{4}(I_L + I_L^{*} - I_R - I_R^{*}) =
\frac{1}{4\pi} \int_{0}^{\infty} d\omega\,\, \omega \qquad\qquad\nonumber \\
{\rm Tr}\Bigl\{(G^r-G^a)(\Sigma_R^{<}-\Sigma_L^{<}) + i G^{<}
(\Gamma_R-\Gamma_L) \Bigr\},\quad \label{eq-current-2}
\end{eqnarray}
where $\Gamma_\alpha = i(\Sigma_\alpha^r - \Sigma_\alpha^a)$.
Equation~(\ref{eq-heat-current}) or (\ref{eq-current-2}) is valid for any
finite temperature differences of the leads.

We define the thermal conductance as
\begin{equation}
\sigma = \lim_{\Delta T \to 0} \frac{I}{\Delta T} = 
\frac{\delta I}{\delta T},
\label{conductance-def}
\end{equation}
where $\Delta T$ is the difference of the temperatures between the leads, such
that $T_L = T + \Delta T/2$ and $T_R = T - \Delta T/2$.  To take the limit
$\Delta T \to 0$ explicitly, we introduce variational derivatives of the
Green's functions, self-energies, or current, in the sense, e.g.,
\begin{equation}
 \frac{\delta G}{\delta T} = \lim_{\Delta T \to 0 }
\frac{G(T_L, T_R) - G(T, T)}{T_L - T_R}.
\end{equation}
This is different from varying the temperature of the whole system.  The
variational derivation measures the response of the system with a fixed
overall temperature $T$ and varying only the temperatures of the leads with
respective to an average temperature.  With the variation derivative defined,
we can express the conductance in a form similar to the Landauer formula
(\ref{eq-landauer}) for the ballistic transport as \cite{jswpre07,nzeng07},
\begin{equation}
\label{eq-conductance-2}
\sigma = 
\frac{1}{2\pi} \int_0^\infty \!\!\!d\omega\, \omega\, \tilde{T}[\omega] 
\frac{\partial f(\omega)}{\partial T}
\end{equation}
with an effective transmission coefficient,
\begin{eqnarray}
\tilde{T}[\omega] &=& 
\frac{1}{2} {\rm Tr} \bigl\{ G^r (\Gamma_L + \frac{1}{2}\Gamma_n - S) G^a \Gamma_R \bigr\} +\nonumber \\
\label{eq-eff-trans}
&&\frac{1}{2} {\rm Tr} \bigl\{ G^a \Gamma_L G^r (\Gamma_R + \frac{1}{2}\Gamma_n +S) \bigr\},
\end{eqnarray}
where the nonlinear effect is reflected in the extra terms,
$\Gamma_n = i (\Sigma_n^r - \Sigma_n^a)$ and 
\begin{equation}
S = \left[ f \frac{\delta \Gamma_n}{\delta T} - 
i\frac{\delta \Sigma_n^{<}}{\delta T} \right]
\left({\partial f \over\partial T}\right)^{-1}.
\end{equation}
When $\Gamma_n = S = 0$, we recover the Caroli formula. We note that the
effective transmission $\tilde T[\omega]$ is temperature dependent, while the
ballistic transmission is not.

\subsubsection{Perturbative and mean-field self-energies}

\begin{figure}
\includegraphics[width=\columnwidth]{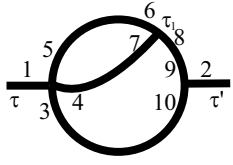}
\caption{\label{fig-feynman13}The 13th Feynman diagram for the nonlinear
self-energy $\Sigma_n$ with labelling.}
\end{figure}

The nonlinear self-energy, $\Sigma_n(\tau, \tau')$, is the key quantity in
solving the heat transport problem in interacting systems.  Clearly, it is
impossible to evaluate it exactly.  Some form of approximation is unavoidable.
As the nonlinear interactions are relatively weak for real systems,
perturbative treatment can be reliable for temperatures not too high.  The
approximation can be improved, at least in principle, if more diagrams can be
included.

The diagrams shown in Fig.~\ref{fig-feynmant4} are for the contour-ordered
functions.  The rules of Feynman diagrams to formula are as follows: label
each leg of a vertex with $(j_l, \tau_l)$ where $l$ runs over the legs of all
the vertices.  A vertex is associated with $T_{j_l, j_{l'}, j_{l''}}(\tau_l,
\tau_{l'}, \tau_{l''})$ for a three-body interaction, and similarly for a
four-body interaction.  If leg $l$ and $l'$ are connected by a line, associate
them with the Green's function $G_{j_l,j_{l'}}^{0}(\tau_l,\tau_{l'})$.
Integrate on the contour all the internal $\tau$ variables and sum over the
internal site indices $j$.  Since each $n$-leg vertex has $n-1$ generalized
delta functions in $\tau$, this effectively makes each vertex with one single
$\tau$.  For example, the 13-th diagram for the self-energy is labelled as
shown in Fig.~\ref{fig-feynman13}, then the associated expression is
\begin{eqnarray}
\sum_{j_3, j_4, j_5, j_6,\atop j_7, j_8, j_9, j_{10}} \int d\tau_1 
T_{j_1,j_3,j_4,j_5} T_{j_6,j_7,j_8} T_{j_9,j_{10}, j_2} 
G_{j_3,j_{10}}^0(\tau,\tau') \nonumber \\
G_{j_5,j_6}^0(\tau,\tau_1) 
G_{j_4,j_7}^0(\tau, \tau_1) G_{j_8,j_9}^0(\tau_1, \tau').\qquad
\end{eqnarray}

High order perturbative treatment is computationally too costly except for
very small systems.  Mean-field approximations are usually used. In such
approximations, only certain selected diagrams are used, with the replacement
of $G^0$ by the full Green's function $G$.  So the self-energy is considered a
functional of $G$ instead $G^0$.  For Coulomb-type interactions in electronic
transport, $GW$ or even simpler treatments have been used
\cite{thygesen07,thygesen-0710.0482}. Certain types of self-consistent
approximation have the added advantage that the energy conservation
\cite{baym62,leeuwen06} is fulfilled in the sense $I_L + I_R = 0$.  This may
not be the case if bare self-energies are used.

\begin{figure}
\includegraphics[width=\columnwidth]{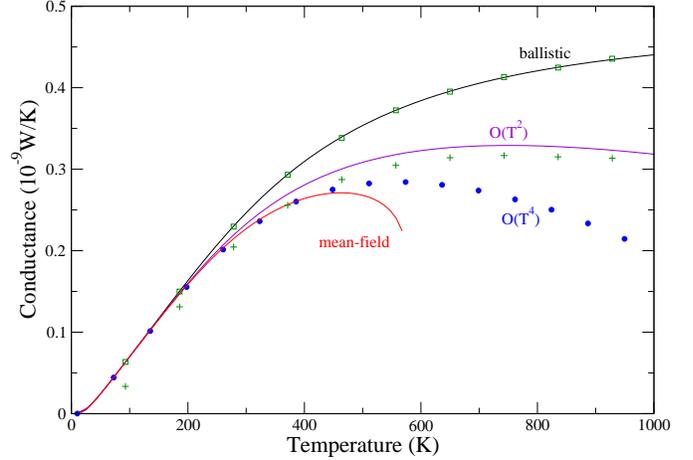}
\caption{\label{fig-kappa1D}Thermal conductance as a function of temperature
for a 1D junction with three atoms in different approximations.  The harmonic
spring constants are $k_L = 1.56$, $k_R = 1.44$, $k_C=1.38$ (eV/(\AA$^2$amu)).
The coupling between the center and leads is the same as the spring constants
of the respective leads.  The nonlinear strength is $t=1.8$
eV/(\AA$^3$amu$^{3/2}$), with an interaction of the form $(t/3)\sum (u_j -
u_{j+1})^3$.  Small onsite quadratic potentials are applied to the leads with
spring constants $k_L^{\rm onsite} = 0.006$, and $k_{R}^{\rm onsite} = 0.013$
(eV/(\AA$^2$amu)).  From ref.~\protect{\cite{jswprb06}}.  The squares and
crosses are the molecular dynamics results for the ballistic and nonlinear
systems, respectively, from this work.  The MD step size is $\Delta t =
3\times 10^{-16}\,$s with $5\times 10^8$ MD steps.}
\end{figure}

In ref.~\cite{mingo06}, only the first self-energy diagram in
Fig.~\ref{fig-feynmant4} is used. We have made a comparison of the first and
second order perturbative approximation, and self-consistent mean-field
approximation (with the first two diagrams), see Fig.~\ref{fig-kappa1D}.  Our
experience with several models \cite{jswprb06,jswpre07} seems to suggest that
mean-field approximation tends to over estimate the nonlinear effect.
Additional problem associated with a self-consistent approximation is that the
iterations may not converge.  Thus, finding a good approximation is still an
open challenge for the phonon interactions at high temperatures.

\subsection{Classical molecular dynamics with quantum heat-baths}

Molecular dynamics (MD) has been used extensively to study thermal transport
based on Green-Kubo formula in equilibrium or for nonequilibrium MD with heat
source and sink (heat baths).  However, since MD follows a purely classical
dynamics, quantum effects are not taken into account. There have been some
attempts to partially incorporate quantum feature into an MD by scaling the
temperature and the thermal conductivity after the simulations
\cite{czwang90,yhlee91,juli98}.  Given the actual classical simulation
temperature $T_{\rm MD}$ with kinetic energy $N k_B T_{\rm MD}/2$, where $N$
is the number of degrees of freedom, a new temperature $T$ can be defined by
comparing to a quantum harmonic system with the same kinetic energy,
\begin{equation}
\label{eq-T-scale}
N k_B T_{\rm MD} = \sum_{k} \hbar \omega_k \left( 
\frac{1}{\exp(\hbar \omega_k/(k_B T)) - 1} 
+ \frac{1}{2}\right),
\end{equation}
where the summation is over the vibrational modes.  The conductivity is also
scaled by \cite{yhlee91}
\begin{equation}
\label{eq-kappa-scale}
\kappa = \kappa_{\rm MD} { d T_{\rm MD} \over dT},
\end{equation}
which is effectively multiplying the classical value by the quantum heat
capacity of the corresponding harmonic lattice.  Both Eq.~(\ref{eq-T-scale})
and (\ref{eq-kappa-scale}) are plausible but lack a more rigorous basis.
Another approach is the centroid molecular dynamics proposed by Cao and Voth
\cite{jscao93,yonetani04}.  This appears very sophisticated. The main
difficulty here is to obtain the temperature-dependent effective potential for
the molecular dynamics.

In ref.~\cite{jswang07}, we proposed a method based on a generalized Langevin
dynamics.  The quantum Langevin dynamics is well studied
\cite{lindenberg90,haanggi05}.  However, its use in thermal transport is only
recent \cite{dharprb06,dhar06}.  The idea is to consider a quantum Langevin
equation where the degrees of freedom of the leads are eliminated in favoring
that of only the central interacting region.  This can be performed exactly if
the leads and the couplings between the center and leads are linear.  Then the
central part is treated as a classical system, while the leads retain
quantum-mechanical properties.  This amounts to a quasi-classical
approximation of the quantum Langevin equation \cite{schmid82,weiss99}.  Such
a treatment also works for electron transport, and electron-phonon
interactions \cite{jtlu08}.

With the Hamiltonian defined by Eq.~(\ref{eq-H1}), the corresponding
quasi-classical generalized Langevin equation is given by \cite{jswang07}
\begin{equation}
\ddot{u}^C = -K^C u^C + F_n(u^C\!)\! - \!\!\int_{t_0}^t\!\! \Sigma^r(t, t')u^C(t') dt' 
+ \xi_L + \xi_R,
\label{eq-langevin}
\end{equation}
where $F_n = - \partial V_n/\partial u^C$ is the nonlinear force, $\Sigma^r$
is the retarded self-energy of the leads, $\Sigma^r = \Sigma_L^r +\Sigma_R^r$,
as used in the NEGF calculation, but in the time domain; $\Sigma_L^r = V^{CL}
g^r_L V^{LC}$.  A similar equation holds for the right lead $\Sigma^r_R$ using
the right lead surface Green's function.

The noise $\xi_L(t)$ or $\xi_R(t)$ has zero mean and a correlation matrix
determined by the temperature and the model of the leads.  Using the (surface)
density of states, the expression for the correlation can be simplified to get
a rather compact result for the spectrum of the noises \cite{dhar06},
\begin{equation}
\tilde{F}[\omega]=\!\int_{-\infty}^\infty\!\!\!\!
\bigl\langle \xi_L(t) \xi_L^T(0)\bigr\rangle  e^{i\omega t} dt 
= \Bigl(f_L(\omega) + \frac{1}{2}\Bigr)
\hbar \Gamma_L[\omega], \label{eq-spectra}
\end{equation}
where $\Gamma_L[\omega] = i \bigl( \Sigma^r_L[\omega] - \Sigma^a_L[\omega]
\bigr) = -2\, {\rm Im}\, V^{CL} g^r_L[\omega] V^{LC}$, and $f_L(\omega)$ is
the Bose-Einstein distribution function at the temperature of left lead.  The
situation for the right lead is analogous.  The spectrum function
$\tilde{F}[\omega]$ is even in $\omega$ and is a symmetric, positive
semidefinite matrix.  A classical limit is obtained if we take
$\bigl(f_L(\omega) + 1/2\bigr)\hbar \approx k_B T_L/\omega$.
 
The steady state thermal current can be computed in several equivalent ways:
\begin{eqnarray}
I_L = -I_R = - \bigl\langle {d H_L \over dt} \bigr\rangle = 
\langle (\dot{u}^L)^T V^{LC} u^C \rangle  \qquad\qquad\nonumber \\
\label{eqIdef}
=-\langle u^C(t)^T \dot{B}(t) \rangle =
 \langle \dot{u}^C(t)^T B(t) \rangle, \quad
\end{eqnarray}
where $B(t) = -\int_{t_0}^t \Sigma^r_L(t,t') u^C(t') dt' + \xi_L(t)$.

\begin{figure}
\includegraphics[width=\columnwidth]{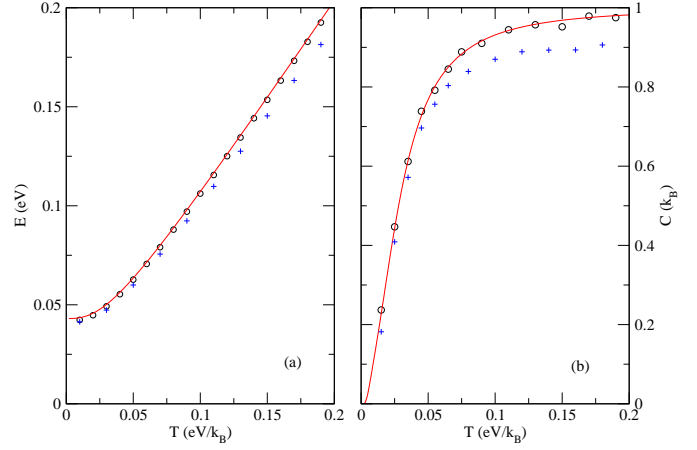}
\caption{\label{fig-eng_cap}Energy (a) and heat capacity (b) per degree of
freedom for the 1D chain models studied in ref.~\protect{\cite{jswang07}}. The
solid lines are exact results for a 1D harmonic chain of length $N=128$ with
periodic boundary condition, while the circles are obtained for the same model
from the MD with quantum heat-baths at the ends; the crosses are the MD
results of a nonlinear model with an extra quartic onsite potential,
$\frac{\mu}{4} \sum u^4_j$, with $\mu=1\,$eV/(\AA$^4$amu$^2$).}
\end{figure}

If $V_n=0$ the linear Langevin equation can be solved analytically.  The
result for the thermal current reproduces exactly the Caroli formula
\cite{dhar06}.  On the other hand, if temperature is sufficiently high (e.g.,
$1/2$ of the Debye temperature), the nonlinear Langevin equation agrees with
classical MD result (with appropriate heat baths).  For nonlinear systems with
$V_n \neq 0$, the equation can be solved numerically.  At low temperatures,
the transport is mostly ballistic, the dynamics correctly reproduces quantum
results and agrees well with NEGF calculation \cite{jswang07}.  At the
intermediate temperature range, where both quantum and nonlinear effects are
strong, the dynamics gives only approximate results.  However, we note that
the quasi-classical approximation can be systematically improved, if we use
the techniques proposed in
\cite{balletine98,prezhdo00,pahl02,heatwole05,prezhdo06,barik03,barik04,horsfield04b}.
For example, the leading order correction is obtained if we enlarge the
equation set to include variables $u^2$, $up$, and $p^2$ ($p=\dot{u}$) as
independent variables. Thus, the dynamics can provide a nonperturbative
solution to the quantum transport problems.

In Fig.~\ref{fig-kappa1D} we show comparisons of MD and NEGF results. For the
ballistic case, the agreement is perfect with very high precision.  For the
nonlinear system, in order to have a stable MD, we have added an additional
quartic potential, $\mu/4 \sum(u_j - u_{j+1})^4$, with $\mu=
0.6\,$eV/(\AA$^4$amu$^2$) in the simulation, thus the comparison is hardly on an
equal footing. The discrepancy at low temperatures between NEGF and the MD has
to do with the zero-point motion contribution.  In the noise spectrum,
Eq.~(\ref{eq-spectra}), the $1/2$ in $f_L + 1/2$ represents a contribution
from zero-point vibration of the quantum oscillators.  We can replace $f_L +
1/2$ by $f_L(\omega)$ for $\omega > 0$ and $-f_L(-\omega)$ for $\omega < 0$,
and it will not have any effect to the thermal current of a ballistic system.
But the removal of the zero-point motions does have some consequences to the
nonlinear system under quasi-class approximation.

It is interesting to note that the Langevin dynamics with quantum heat baths
also provides a way to compute the equilibrium energy, and by numerical
differentiation, the heat capacity of the quantum models. The energy is
obtained by the MD averages of the classical expression of kinetic energy plus
the potential energy, with a left and right bath at the same temperature. A
comparison of the exact and MD results is presented in Fig.~\ref{fig-eng_cap}
for the linear chain. For nonlinear systems in thermal equilibrium, an
alternative method is (path-integral) quantum Monte Carlo \cite{landau05}.
 
\subsection{Other treatments of nonlinear effects}
In this subsection, we briefly mention several other approaches for treating
the nonlinear quantum thermal transport for dielectric materials. In
refs.~\cite{michel05,michel06,michelrev06} Michel {\sl et al.} used Hilbert
space average method to derive diffusion-like equations on 1D chains, thus
justified the Fourier's law and computed the thermal conductivity from
Hamiltonian model parameters. These models are idealized so that each site has
a band of excited states separated with an energy gap to the ground state.
The couplings between the sites are assumed weak.  Diffusion equation of the
form, $dP_{\mu}/dt = \gamma (P_{\mu-1}- 2P_{\mu} + P_{\mu+1})$, can be
derived, where $P_{\mu}$ is the probability that site $\mu$ is at excited
states.  The diffusion constant $\gamma$ is then related to the thermal
conductivity.  Segal {\sl et al.} studied thermal rectifying effects of
quantum systems with somewhat similar approach of subspace projection, where
again diffusion-like rate equations are derived for the occupation of the
quantum states \cite{segal05,segaljcp05,segal06}, the nonlinearity is
introduced by restricting to two-level or finite-level systems.  In
ref.~\cite{segal-arxiv07} a master equation approach for more general systems
is considered and the Fourier's law of heat conduction is derived.  The
approximations made always put the systems in diffusive regime.  It seems to
us that it is difficult to address the issue of transition from ballistic
behavior at short time and length scales to diffusive behavior at long time
and length scales in the above approaches, but see ref.~\cite{steinigeweg07}
for such possibilities.  In refs.~\cite{buldum99,leitner00,leitner01} heat
flow through model molecular and nanocrystal junctions is treated using
Fermion-golden rule for the rate of phonon relaxations.  The Fermi-Pasta-Ulam
chain is treated quantum-mechanically in ref.~\cite{santhosh07}. The authors
find an $L^{0.4}$ divergence for the thermal conductivity by considering the
relaxation rates of the phonon modes.

\section{Electron-phonon interactions}
In this section, we extend the NEGF method in subsec.~\ref{sec-NEGF} to
include the electron subsystem and electron-phonon interactions (EPIs). The
study of EPI in the context of nanoscale electronic transport is an important
field which has attracted intense research recently
\cite{galperin-review07,mitra04}.  An extensive discussion of this issue is
certainly out of the scope of current review. Here we only concentrate on one
special aspect, namely the current induced heating in nanostructures.
Different schemes exist in the literature to study this effect
\cite{horsfield04,todorov98,agrait,segal:3915,montgomery02,horsfield05,horsfield06,ycchen03,frederiksen04,pecchia07,sun07,lu07}.
When the EPI is strong, a canonical transformation is useful to study a
minimum model system \cite{mahan00,galperin-review07,galperin07,sun07}. But it
is not applicable to large systems.  On the other hand, perturbative
approaches based on the lowest order Born approximation are applicable for
relatively weak interactions \cite{todorov98,pecchia07,galperin07,lu07}.
Actually, it has already been used in the first principles study
\cite{ycchen03,frederiksen04,pecchia07}. Hybrid approaches also exist, where
the electrons are treated quantum-mechanically, while the phonons use MD with
quantum corrections
\cite{horsfield04,montgomery02,horsfield05,horsfield06}. In our NEGF
formalism, we use the mean field method based on the lowest order Born
approximation, the so-called self-consistent Born approximation (SCBA).

The Hamiltonian of the whole system is the sum of the electron and phonon
part.  Consisting with Eq.~(\ref{eq-H1}), the electron Hamiltonian reads
\begin{equation}
\label{eh0}
{\cal H^{\rm e}} = \!\!\!\!\!\sum_{\alpha=L,C,R}\!\!\!\!\!{c_\alpha^\dagger} F^\alpha c_\alpha  + \sum_{\alpha=L,R}\!\!\!\!\!\left(c^\dagger_\alpha V_{\rm e}^{\alpha C} c_C + {\rm h.c.}\right) +  \sum_{ijk}M_{ij}^ku_kc_i^\dagger c_j,
\end{equation}
where $c_\alpha$ is a column vector consisting of all the annihilation
operators in $\alpha$ region, $c_\alpha^\dagger$ is the corresponding creation
operators. $V_{\rm e}^{\alpha C}$ has similar meaning as $V^{\alpha C}$ in the
phonon Hamiltonian, and $V_e^{\alpha C} = (V_e^{C\alpha})^\dagger$. The last
term represents the EPI, where we only keep the first order Taylor expansion
about the atomic equilibrium position. $M_{ij}^k$ is the interaction
coefficient. We assume that EPI takes place only at the center region, and the
subscript $C$ has been omitted.

In ref.~\cite{lu07} we adopt the SCBA to study the coupled electron and phonon
transport. All the electron Green's functions are defined as usual
\cite{haug96}, e.g.,
\begin{equation}
	D^r(t,t') = -i\theta(t-t')\langle[c(t),c^\dagger(t')]_+\rangle.
	\label{eq:ed1}
\end{equation}
We use $d$ to denote the free lead Green's function. $D_0$ and $D$ are the
noninteracting and the full electron Green's functions of the center,
respectively. The free lead Green's function $d$ can be obtained using the
same method as that for phonons ($\alpha=L, R$)
\begin{equation}
	d^{r,a}_\alpha[\omega] = \bigl[(\omega\pm i\eta)-F^\alpha\bigr]^{-1},
	\label{eq:el1}
\end{equation}
\begin{equation}
	d^<_\alpha[\omega] = -f^{\rm e}_\alpha[\omega]\Bigl(d^r_\alpha[\omega]-d^a_\alpha[\omega] \Bigr).
	\label{eq:el3}
\end{equation}
$f^{\rm e}(\omega)$ is the Fermi-Dirac distribution function. The
noninteracting electron Green's functions can also be solved exactly
\begin{equation}
	D_0^{r,a}[\omega] = \left[(\omega\pm i\eta)-F^C-\Pi^{r,a}[\omega]\right]^{-1},
	\label{eq:el5}
\end{equation}
\begin{equation}
	D_0^{<,>}[\omega] = D_0^r[\omega]\Pi^{<,>}[\omega]D_0^a[\omega].
	\label{eq:el6}
\end{equation}
The electron self-energy includes contributions from both leads $\Pi =
\Pi_L+\Pi_R$, and
\begin{equation}
	\Pi_\alpha^{r,a,<,>}[\omega] = V_{\rm e}^{C\alpha}d^{r,a,<,>}_\alpha[\omega]V_{\rm e}^{\alpha C}.
	\label{eq:el7}
\end{equation}
The full Green's function $D$ follows the same relations, except that the
self-energies include EPI contribution. The notations used in
Eq.~(\ref{eq:ed1}) to (\ref{eq:el7}) for electron Green's functions and
self-energies are not standard.  As we have already used $G$ and $\Sigma$ for
the phonons in earlier sections, $D$ and $\Pi$ are used here for the
electrons.

The central problem is how to get $\Pi_{\rm epi}$ and $\Sigma_{\rm epi}$.  The
SCBA includes only the lowest order self-energies. For $\Pi_{\rm epi}$, there
are two Feynman diagrams, the first two in Fig.~\ref{fig-feynmant4} with
modifications. In the first one, only half of the circle represents phonons,
and all other lines represent electrons. In the second one, only the vertical
line represents phonons, and all others are electrons. For $\Sigma_{\rm epi}$,
there is only one diagram, the first one in Fig.~\ref{fig-feynmant4} with the
circle representing electrons. The coefficients also change to $i$, $-i$, and
$-i$, respectively.  The rules to translate the diagrams into expressions
remain the same as for phonons, provided that the interaction vertex is
defined by
\begin{equation}
	M^k_{ij}(\tau_i,\tau_j,\tau_k) =
M^k_{ij}\delta(\tau_i,\tau_j)\delta(\tau_i,\tau_k).
	\label{eq:eph-vertex}
\end{equation}
Mean field method is used to calculate the self-energies and the Green's
functions.

The electron and the energy current can be expressed via the Green's
functions. The electronic current out of the lead $\alpha$ ($\alpha = L,R$) is
\cite{meir92,jauho94,haug96}
\begin{equation}
	J_\alpha = e \int_{-\infty}^{+\infty}\frac{d\omega}{2\pi}\, \mathrm{Tr}\bigl\{D^>[\omega] \Pi^<_\alpha[\omega] - D^<[\omega]\Pi^>_\alpha[\omega]\bigr\}.
	\label{eq:ecur}
\end{equation}
Similarly, the energy current is
\begin{equation}
	I^{\rm e}_\alpha = \int_{-\infty}^{+\infty} \frac{d\omega}{2\pi} \,\omega \mathrm{Tr}\bigl\{D^>[\omega] \Pi^<_\alpha[\omega] - D^<[\omega]\Pi^>_\alpha[\omega]\bigr\}.
	\label{eq:eecur}
\end{equation}
The heat current can be obtained from Eqs.~(\ref{eq:ecur}--\ref{eq:eecur}) as
$I^{h}_\alpha = I^{\rm e}_\alpha-\mu_\alpha J_\alpha/e$. $\mu_\alpha$ is the
chemical potential of the lead $\alpha$.  For phonons the energy and the heat
current are the same. In the presence of EPI, there is energy exchange between
the electron and the phonon subsystem, which is the source of Joule
heating. The use of SCBA guarantees the energy and electronic current
conservation \cite{lu07,thygesen-0710.0482}, so that we can write them in
symmetric forms.  The electronic current is
\begin{equation}
\label{eq:ecur1}
J = e \int_{-\infty}^{+\infty} \frac{d\omega}{2\pi}\, \tilde{T}^e[\omega]\Bigl(f^e_L[\omega]-f^e_R[\omega]\Bigr).
\end{equation}
The effective transmission coefficient reads ($\omega$ omitted)
\begin{eqnarray}
\label{eq:etransmission}
\tilde{T}^e = & \mathrm{Tr}\Bigl\{\frac{1}{2}\bigl [D^r(\Lambda^e_L+\frac{1}{2}\Lambda_{\rm epi}-X_{\rm epi})D^a\Lambda^{\rm e}_R \bigr.\Bigr.\nonumber \\
& \Bigl. \bigl. +D^a\Lambda_L^{\rm e}D^r(\Lambda^{\rm e}_R+\frac{1}{2}\Lambda_{\rm epi} + X_{\rm epi})\bigr]\Bigr\},
\end{eqnarray}
\begin{equation}
\label{eq:etransmission2}
X_{\rm epi} = \frac{\frac{1}{2}(f^e_R+f^e_L)\Lambda_{\rm epi} +
i\Pi_{\rm epi}^<}{f^e_L-f^e_R},  
\end{equation}
$\Lambda^{\rm e}_\alpha=i(\Pi^r_\alpha-\Pi^a_\alpha)$ ($\alpha=L,R$) and $\Lambda_{\rm epi} =
i(\Pi^r_{\rm epi}-\Pi^a_{\rm epi})$.
The total energy current includes electron and phonon contributions
\begin{eqnarray}
\label{eq:etcur}
I^{\rm t} = & \displaystyle\int_{-\infty}^{+\infty} {d\omega \over 2\pi}\, \omega \Big\{\tilde{T}^{\rm e}[\omega]\bigl(f^{\rm e}_L(\omega)-f^{\rm e}_R(\omega)\bigr)\nonumber\\
&+\frac{1}{2}\tilde{T}^{\rm ph}[\omega]\bigl(f_L(\omega)-f_R(\omega)\bigr)\Big\},
\end{eqnarray}
where $\tilde{T}^{\rm ph}$ is similar to eq.~(\ref{eq-eff-trans}), except that
$\Gamma_n$ is replaced by $\Gamma_{\rm epi} = i(\Sigma_{\rm epi}^r-\Sigma_{\rm
epi}^a)$, and $S$ by
\begin{equation}
	S_{\rm epi} = \frac{\frac{1}{2}(f_R+f_L)\Gamma_{\rm epi} - i\Sigma_{\rm epi}^<}{f_L-f_R}.
	\label{eq:phs}
\end{equation}
Here we have corrected a sign error in Eq.~(22) of ref.~\cite{lu07}.  Heat
generation during the current flow can also be written in terms of Green's
functions \cite{lu07},
\begin{equation}
	\label{eq:heatgen}
Q = i\!\int\limits_{-\infty}^{+\infty}\!\! \frac{d\omega'}{2\pi} \int\limits_{-\infty}^{+\infty}\!\! \frac{d\omega}{2\pi} \omega\!\!\!
\sum_{nmikl}\!\!\! D^>_{nm}[\omega']M_{mi}^kG_{kl}^<[\omega]D^<_{ij}[\omega'\!\!-\!\omega]M_{jn}^l.
\end{equation}
Equation~(\ref{eq:heatgen}) can be used to study the Joule heating in
nonequilibrium molecular junctions. Similar results have also been derived by
other authors \cite{galperin07,pecchia07,sun07}.

\begin{figure}
\includegraphics[width=\columnwidth]{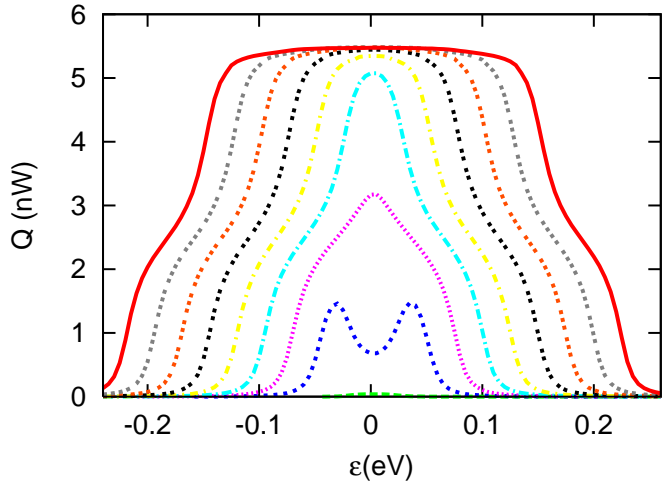}
\caption{\label{fig-heating}Heat generation $Q$ as a function of electrical
onsite energy $\varepsilon$ for a quantum dot coupled with 1D leads
(ref. \cite{lu07}). From the inner to the outer side, the applied biases are
$V = 0.05$ to $0.50$ V with an increment of 0.05 V, respectively.}
\end{figure}

In Fig.  \ref{fig-heating}, we show the heat generation as a function of
onsite energy of a single quantum dot coupled with two 1D leads, calculated
using Eq.~(\ref{eq:heatgen}).  When the applied bias is less than the phonon
energy $\omega$, there is no heating.  When the bias energy is slightly larger
than $\omega$ and less than $2\omega$, there are two energies where the heat
generation is the largest.  These two peak positions are approximately at
$-0.5\,{\rm eV}+\omega$ and $0.5\,{\rm eV}-\omega$.  They merge into a single
one at a bias of $2\omega$ until it reaches saturation. After that,
this peak broadens to a ladder.  These interesting features are unique
properties of nanostructures, absent in the bulk case.
 
\section{Experiments in nanostructure heat transport}

Some of the earliest direct mesoscopic thermal conductance measurements are
done on patterned semiconductor structures with cross section sizes of few
hundred nanometers \cite{tighe97}. The calorimetry on such a small scale is a
challenge, but very promising, e.g., one may count the number of phonons in a
nanostructure \cite{roukes99}. In 2000, such type of devices is used to
confirm the universal thermal conductance quanta at temperatures below $1\,$K
\cite{schwab00}. Bridges of catenoidal shape are used, based on an earlier
theoretical analysis that such geometric shape gives the maximum transmission
\cite{rego98}.  Other extensively studied systems are carbon nanotubes.
Initially, only mats or rope bundle or film form can be measured, given a
relatively low thermal conductivity of order 100 W/(mK) at room temperature
\cite{hone99,djyang02,hqxie07}.  The initial theoretical predictions by
molecular dynamics for single nanotubes along the axial direction are much
higher \cite{berber00,lukes07}.  Indeed, individual multiwalled and single
wall carbon nanotube measurements became possible, and the measured values are
about 3000 W/(mK) at room temperature by several groups
\cite{pkim02,pkim02b,cyu05,pop06}.  Typically, these measurements use tube
lengths of few $\mu$m.  The length dependence effects are analyzed in
\cite{pop06} and observed in \cite{zlwang07}.  In ref.~\cite{hychiu05}, it is
claimed that ballistic phonon transport is consistent with experimental data,
even for temperatures as high as $900\,$K for submicron tube lengths.  Another
class of systems is individual silicon nanowires
\cite{dyli03,bourgeois07,hochbaum08}.  It has been observed that rough and
small diameter silicon nanowires have a low thermal conductivity; this
property can be used for good thermoelectric applications \cite{hochbaum08}.
Thermal rectification effect has been observed in carbon nanotubes with one
side of the tube deposited with C$_9$H$_{16}$Pt \cite{cwchang06}.  However,
the observed effect is rather small, at a level of few percents. Another way
of explicit control of the heat flow in nanostructure is using a tunable
thermal links \cite{cwchang07}, where a multiwalled carbon nanotube is cut and
the outer layers of the tube can be moved mechanically with respect to the
inner layers.  Much large modulation is then possible.  In
ref.~\cite{cwwang07prl}, it is demonstrated that the carbon nanotube heat
conduction is rather robust again bending and deformation.  Phonon transport
in point contact of metal-insulator is studied in ref.~\cite{feher07}. The
studies of thermal transport at an individual molecular level are scarce.  But
impressive progress is made recently \cite{zbge06,nitzan07,zhwang07}.  In
ref.~\cite{zhwang07}, alkane chain anchored to a gold substrate is heated to
about $1100\,$K, the propagation of the heat waves is timed.  This timing
information indicates that the heat transport at such scale of few nanometers
is ballistic.  The thermal conductance was inferred to be $0.05\,$nW/K.

\section{Conclusion}
Using a general junction model as an illustration, we consider the problem of
theoretical modeling of heat transport in nanojunctions.  If the nonlinear
forces can be omitted, this would be the case if temperature is low or the
sizes of the system is small, the Landauer formula gives a very good
description of the ballistic thermal transport.  We give a derivation of the
Landauer formula from the wave scattering point of view.  A derivation from
NEGF is also outlined as a special case of the more general nonlinear result.
Several different ways of calculating the transmission coefficient appearing
in the Landauer formula are reviewed.  Among the alternative computational
methods, the calculation based on Caroli formula and the iterative algorithm
of the surface Green's functions is the most efficient computationally for
the total transmission coefficient $T[\omega]$.  The mode-matching method
gives the results of transmission of each mode to every other mode, thus
containing more information. The relation of the two approaches is clarified.
The transfer matrix method and Piccard formula are theoretically interesting,
but they may not be practical in actual calculations, e.g., the transfer
matrix can be numerically unstable for large systems.  In summary, for the
ballistic transport, efficient computational algorithms exist for arbitrary
complex systems with numbers of atoms less than about $10^3$.

For the treatment of nonlinear effects, it is still a great challenge.  We
first presented a phenomenological treatment of the nonlinear effect using the
information on the mean free path.  A systematic development of the
nonequilibrium Green's function method is given.  We emphasize the use of
contour ordered Green's functions and the equations of motion defined on the
Keldysh contour.  To do this, we must generalize the $\theta$ and $\delta$
functions, and also generalize the Heisenberg equation to the contour.
Quantum field-theoretic method is used to develop the Dyson equations relating
the ballistic system and interacting system.  The nonlinear effect can be
taken into account if the nonlinear self energy $\Sigma_n$ can be obtained
accurately.  Perturbative expansion and mean-field approximation to the
nonlinear self-energy are possible.  The Landauer formula can be generalized
formally in the nonlinear case, except now the transmission function is
temperature dependent.  For small systems at not too high temperatures
(e.g., $T < 400\,$K), the results with these approximations are reliable.  But
for large systems or higher temperatures, the approximation can break down.  A
more promising method to handle large systems and high temperatures is
molecular dynamics.  With the quantum heat baths discussed in this review, the
quantum effect can be partially taken into account and it is even possible 
to correct the quasi-classical approximation.

We also reviewed briefly methods for electron-phonon interactions.  This is an
important area where many interesting effects can be investigated
experimentally, and it is also very relevant to the electronic industry.  NEGF
and other methods will play important role in the study.  We also give a quick
review of the experimental status of thermal transport problems in
nanostructures.  To a theoretical or computational modelling scientist, it is
easy to work with small systems, while it is hard to an experimentalist to
manipulate and measure heat at the molecular level.  We hope that the two
sides of approaches will converge well in the near future.

For the convenience of reader, we have put up a webpage \cite{negfweb} to post
some of the implementations of algorithms, which may clarify the doubts and
subtlties in understanding them.

\section*{Acknowledgments}
Parts of the results under review were in collaboration with Nan Zeng and
Chee-Kwan Gan.  The authors thanks for discussions with Baowen Li, Jiangbin
Gong, Pawel Keblinski, Saikong Chin, Yong Xu, and Jinhua Lan.  This work is
supported in part by a faculty research grant R-144-000-173-112/101 of NUS.

%
%
%
%
\bibliographystyle{epj}
\bibliography{transp}

\begin{thebibliography}{218}

\bibitem{peierls55}
R.E. Peierls, \emph{Quantum Theory of Solids} (Oxford Univ., 1955)

\bibitem{carruthers61}
P.~Carruthers, Rev. Mod. Phys. \textbf{33}, 92 (1961)

\bibitem{spohn06}
H.~Spohn, J. Stat. Phys. \textbf{124}, 1041 (2006)

\bibitem{cahill03}
D.G. Cahill, W.K. Ford, K.E. Goodson, G.D. Mahan, A.~Majumdar, H.J. Maris,
  R.~Merlin, S.R. Phillpot, J. Appl. Phys. \textbf{93}, 793 (2003)

\bibitem{gchen04}
G.~Chen, D.~{Borca-Tasciuc}, R.G. Yang, in \emph{Encyclopedia of Nanoscience
  and Nanotechnology}, edited by H.S. Nalwa (Amer. Sci. Publ., 2004), Vol.~7,
  p. 429

\bibitem{galperin-review07}
M.~Galperin, M.A. Ratner, A.~Nitzan, J. Phys.:Condens. Matter \textbf{19},
  103201 (2007)

\bibitem{meier69}
P.F. Meier, Phys. kondens. Materie \textbf{8}, 241 (1969)

\bibitem{mahan87}
G.D. Mahan, Phys. Rep. \textbf{145}, 251 (1987)

\bibitem{lepri-review03}
S.~Lepri, R.~Livi, A.~Politi, Phys. Rep. \textbf{377}, 1 (2003)

\bibitem{mcgaughey06}
A.J.H. McGaughey, M.~Kaviany, Adv. in Heat Transf. \textbf{39}, 169 (2006)

\bibitem{heino07}
P.~Heino, J. Comput. Theor. Nanoscience \textbf{4}, 896 (2007)

\bibitem{jswprb06}
J.S. Wang, J.~Wang, N.~Zeng, Phys. Rev. B \textbf{74}, 033408 (2006)

\bibitem{jswpre07}
J.S. Wang, N.~Zeng, J.~Wang, C.K. Gan, Phys. Rev. E \textbf{75}, 061128 (2007)

\bibitem{landauer57}
R.~Landauer, IBM J. Res. Dev. \textbf{1}, 223 (1957)

\bibitem{landauer70}
R.~Landauer, Philos. Mag. \textbf{21}, 863 (1970)

\bibitem{angelescu98}
D.E. Angelescu, M.C. Cross, M.L. Roukes, Superlattices Microsctruc.
  \textbf{23}, 673 (1998)

\bibitem{rego98}
L.G.C. Rego, G.~Kirczenow, Phys. Rev. Lett. \textbf{81}, 232 (1998)

\bibitem{rego01}
L.G.C. Rego, Phys. Stat. Sol. (a) \textbf{187}, 239 (2001)

\bibitem{blencowe99}
M.P. Blencowe, Phys. Rev. B \textbf{59}, 4992 (1999)

\bibitem{blencowe04}
M.~Blencowe, Phys. Rep. \textbf{395}, 159 (2004)

\bibitem{mingo05}
N.~Mingo, D.A. Broido, Phys. Rev. Lett. \textbf{95}, 096105 (2005)

\bibitem{gaussian03}
M.J. Frisch, et~al., gaussian 03, Revision C.02 (Gaussian Inc., Wallingford,
  2004)

\bibitem{wangj05}
J.~Wang, J.S. Wang, cond-mat/0509092

\bibitem{pendry83}
J.B. Pendry, J. Phys. A: Math. Gen. \textbf{16}, 2161 (1983)

\bibitem{blencowe00}
M.P. Blencowe, V.~Vitelli, Phys. Rev. A \textbf{62}, 052104 (2000)

\bibitem{schwab00}
K.~Schwab, E.A. Henriksen, J.M. Worlock, M.L. Roukes, Nature \textbf{404}, 974
  (2000)

\bibitem{chiatti06}
O.~Chiatti, J.T. Nicholls, Y.Y. Proskuryakov, N.~Lumpkin, I.~Farrer, D.A.
  Ritchie, Phys. Rev. Lett. \textbf{97}, 056601 (2006)

\bibitem{meschke06}
M.~Meschke, W.~Guichard, J.P. Pekola, Nature \textbf{444}, 187 (2006)

\bibitem{santamore01}
D.H. Santamore, M.C. Cross, Phys. Rev. B \textbf{63}, 184306 (2001)

\bibitem{santamore01prl}
D.H. Santamore, M.C. Cross, Phys. Rev. Lett. \textbf{87}, 115502 (2001)

\bibitem{cross01}
M.C. Cross, R.~Lifshitz, Phys. Rev. B \textbf{64}, 085324 (2001)

\bibitem{cmchang05}
C.M. Chang, M.R. Geller, Phys. Rev. B \textbf{71}, 125304 (2005)

\bibitem{sxqu04}
S.X. Qu, M.R. Geller, Phys. Rev. B \textbf{70}, 085414 (2004)

\bibitem{cleland01}
A.N. Cleland, D.R. Schmidt, C.S. Yung, Phys. Rev. B \textbf{64}, 172301 (2001)

\bibitem{tanaka04}
Y.~Tanaka, F.~Yoshida, S.~Tamura, Phys. Stat. Sol. (c) \textbf{1}, 2625 (2004)

\bibitem{tanaka05}
Y.~Tanaka, F.~Yoshida, S.~Tamura, Phys. Rev. B \textbf{71}, 205308 (2005)

\bibitem{wxli03}
W.X. Li, K.Q. Chen, W.~Duan, J.~Wu, B.L. Gu, J. Phys. D: Appl. Phys.
  \textbf{36}, 3027 (2003)

\bibitem{wxli04}
W.X. Li, K.Q. Chen, W.~Duan, J.~Wu, B.L. Gu, J. Phys.: Condens. Matter
  \textbf{16}, 5049 (2004)

\bibitem{wxli04apl}
W.X. Li, K.Q. Chen, W.~Duan, J.~Wu, B.L. Gu, Appl. Phys. Lett. \textbf{85}, 822
  (2004)

\bibitem{kqchen05}
K.Q. Chen, W.X. Li, W.~Duan, Z.~Shuai, B.L. Gu, Phys. Rev. B \textbf{72},
  045422 (2005)

\bibitem{wqhuang05}
W.Q. Huang, K.Q. Chen, Z.~Shuai, L.~Wang, W.~Hu, B.S. Zou, J. Appl. Phys.
  \textbf{98}, 093524 (2005)

\bibitem{wxli06}
W.~Li, K.~Chen, Phys. Lett. A \textbf{357}, 378 (2006)

\bibitem{lmtang06}
L.M. Tang, L.L. Wang, K.Q. Chen, W.Q. Huang, B.S. Zou, Appl. Phys. Lett.
  \textbf{88}, 163505 (2006)

\bibitem{wxli06apl}
W.X. Li, T.~Liu, C.~Liu, Appl. Phys. Lett. \textbf{89}, 163104 (2006)

\bibitem{yming06}
Y.~Ming, Z.X. Wang, Z.J. Ding, Phys. Lett. A \textbf{350}, 302 (2006)

\bibitem{xfpeng07}
X.F. Peng, K.Q. Chen, B.S. Zou, Y.~Zhang, Appl. Phys. Lett. \textbf{90}, 193502
  (2007)

\bibitem{pyang07}
P.~Yang, Q.~feng Sun, H.~Guo, B.~Hu, Phys. Rev. B \textbf{75}, 235319 (2007)

\bibitem{jdlu07}
J.D. Lu, L.~Shao, Y.L. Hou, L.~Yi, Chin. Phys. Lett. \textbf{24}, 793 (2007)

\bibitem{ywang07}
Y.~Wang, L.L. Wang, L.M. Tang, B.S. Zou, L.H. Zhao, J. Phys. D: Appl. Phys.
  \textbf{40}, 7159 (2007)

\bibitem{prasher07}
R.~Prasher, T.~Tong, A.~Majumdar, J. Appl. Phys. \textbf{102}, 104312 (2007)

\bibitem{pyang04}
P.~Yang, Ph.D. thesis, McGill Univ. (2004)

\bibitem{wangj06prb}
J.~Wang, J.S. Wang, Phys. Rev. B \textbf{74}, 054303 (2006)

\bibitem{ashcroft-mermin76}
N.W. Ashcroft, N.D. Mermin, \emph{Solid State Physics} (Saunders College, 1976)

\bibitem{hardy63}
R.J. Hardy, Phys. Rev. \textbf{132}, 168 (1963)

\bibitem{khomyakov04}
P.A. Khomyakov, G.~Brocks, Phys. Rev. B \textbf{70}, 195402 (2004)

\bibitem{hzhao05}
H.~Zhao, J.B. Freund, J. Appl. Phys. \textbf{97}, 024903 (2005)

\bibitem{cho05}
K.~Cho, J.~Ushida, M.~Bamba, J. Phys. Soc. Jpn. \textbf{74}, 3088 (2005)

\bibitem{velev04}
J.~Velev, W.~Butler, J. Phys.: Condens. Matter \textbf{16}, R637 (2004)

\bibitem{ando91}
T.~Ando, Phys. Rev. B \textbf{44}, 8017 (1991)

\bibitem{khomyakov05}
P.A. Khomyakov, G.~Brocks, V.~Karpan, M.~Zwierzycki, P.J. Kelly, Phys. Rev. B
  \textbf{72}, 035450 (2005)

\bibitem{NR}
W.H. Press, S.A. Teukolsky, W.T. Vetterling, B.P. Flannery, \emph{Numerical
  Recipes: The Art of Scientific Computing}, 3rd~edn. (Cambridge Univ., 2007)

\bibitem{ting92}
D.Y. Ting, E.T. Yu, T.C. McGill, Phys. Rev. B \textbf{45}, 3583 (1992)

\bibitem{ting99}
D.Y. Ting, Microelec. J. \textbf{30}, 985 (1999)

\bibitem{ptong99}
P.~Tong, B.~Li, B.~Hu, Phys. Rev. B \textbf{59}, 8639 (1999)

\bibitem{macia00}
E.~Maci\'a, Phys. Rev. B \textbf{61}, 6645 (2000)

\bibitem{antonyuk04}
V.B. Antonyuk, M.~Larsson, A.G. Mal'shukov, K.A. Chao, Semicond. Sci. Technol.
  \textbf{20}, 347 (2005)

\bibitem{lscao05}
L.S. Cao, R.W. Peng, R.L. Zhang, X.F. Zhang, M.~Wang, X.Q. Huang, A.~Hu, S.S.
  Jiang, Phys. Rev. B \textbf{72}, 214301 (2005)

\bibitem{ymzhang05}
Y.M. Zhang, S.J. Xiong, Phys. Rev. B \textbf{72}, 115305 (2005)

\bibitem{zhangym07}
Y.M. Zhang, C.H. Xu, S.J. Xiong, Chin. Phys. Lett. \textbf{24}, 1017 (2007)

\bibitem{murphy07}
P.G. Murphy, J.E. Moore, Phys. Rev. B \textbf{76}, 155313 (2007)

\bibitem{hyin07}
H.~Yin, R.~Tao, arXiv:0709.3681

\bibitem{pichard84}
J.L. Pichard, Ph.D. thesis, Univ. of Paris at Orsay (1984)

\bibitem{beenakker97}
C.W.J. Beenakker, Rev. Mod. Phys. \textbf{69}, 731 (1997)

\bibitem{botten07}
L.C. Botten, A.A. Asatryan, N.A. Nicorovici, R.C. McPhedran, C.M. de~Sterke,
  Physica B \textbf{394}, 320 (2007)

\bibitem{caroli71}
C.~Caroli, R.~Combescot, P.~Nozieres, D.~Saint-James, J. Phys. C: Solid St.
  Phys. \textbf{4}, 916 (1971)

\bibitem{meir92}
Y.~Meir, N.S. Wingreen, Phys. Rev. Lett. \textbf{68}, 2512 (1992)

\bibitem{ozpineci01}
A.~Ozpineci, S.~Ciraci, Phys. Rev. B \textbf{63}, 125415 (2001)

\bibitem{segal03}
D.~Segal, A.~Nitzan, P.~H{\"a}nggi, J. Chem. Phys. \textbf{119}, 6840 (2003)

\bibitem{mingo03}
N.~Mingo, L.~Yang, Phys. Rev. B \textbf{68}, 245406 (2003)

\bibitem{wzhang07}
W.~Zhang, T.S. Fisher, N.~Mingo, Numer. Heat Transf. Part B \textbf{51}, 333
  (2007)

\bibitem{yamamoto06}
T.~Yamamoto, K.~Watanabe, Phys. Rev. Lett. \textbf{96}, 255503 (2006)

\bibitem{dhar06}
A.~Dhar, D.~Roy, J. Stat. Phys. \textbf{125}, 805 (2006)

\bibitem{galperin07}
M.~Galperin, A.~Nitzan, M.A. Ratner, Phys. Rev. B \textbf{75}, 155312 (2007)

\bibitem{paulsson02}
M.~Paulsson, cond-mat/0210519

\bibitem{dhlee81}
D.H. Lee, J.D. Joannopoulos, Phys. Rev. B \textbf{23}, 4997 (1981)

\bibitem{guinea83}
F.~Guinea, C.~Tejedor, F.~Flores, E.~Louis, Phys. Rev. B \textbf{28}, 4397
  (1983)

\bibitem{sancho84}
M.P.L. Sancho, J.M.L. Sancho, J.~Rubio, J. Phys. F: Met. Phys. \textbf{14},
  1205 (1984)

\bibitem{sancho85}
M.P.L. Sancho, J.M.L. Sancho, J.~Rubio, J. Phys. F: Met. Phys. \textbf{15}, 851
  (1985)

\bibitem{umerski97}
A.~Umerski, Phys. Rev. B \textbf{55}, 5266 (1997)

\bibitem{sanvito99}
S.~Sanvito, C.J. Lambert, J.H. Jefferson, A.M. Bratkovsky, Phys. Rev. B
  \textbf{59}, 11936 (1999)

\bibitem{krstic02}
P.S. Krsti\'{c}, X.G. Zhang, W.H. Butler, Phys. Rev. B \textbf{66}, 205319
  (2002)

\bibitem{fisher81}
D.S. Fisher, P.A. Lee, Phys. Rev. B \textbf{23}, 6851 (1981)

\bibitem{brenner02}
D.W. Brenner, O.A. Shenderova, J.A. Harrison, S.J. Stuart, B.~Ni, S.B. Sinnott,
  J. Phys.: Condens. Matter \textbf{14}, 783 (2002)

\bibitem{kondo06}
N.~Kondo, T.~Yamamoto, K.~Watanabe, Jpn. J. Appl. Phys. \textbf{45}, L963
  (2006)

\bibitem{yamamoto07}
T.~Yamamoto, Y.~Nakazawa, K.~Watanabe, New J. Phys \textbf{9}, 245 (2007)

\bibitem{schelling02}
P.K. Schelling, S.R. Phillpot, P.~Keblinski, Appl. Phys. Lett. \textbf{80},
  2484 (2002)

\bibitem{becker04}
B.~Becker, P.K. Schelling, S.R. Phillpot, Phys. Stat. Sol. (c) \textbf{1}, 2955
  (2004)

\bibitem{schelling04}
P.K. Schelling, S.R. Phillpot, P.~Keblinsk, J. Appl. Phys. \textbf{95}, 6082
  (2004)

\bibitem{bodapati07}
A.~Bodapati, P.K. Schelling, P.~Keblinski, preprint

\bibitem{yamamotoprl04}
T.~Yamamoto, S.~Watanabe, K.~Watanabe, Phys. Rev. Lett. \textbf{92}, 075502
  (2004)

\bibitem{yamamoto04}
T.~Yamamoto, K.~Watanabe, K.~Mii, Phys. Rev. B \textbf{70}, 245402 (2004)

\bibitem{saito07}
K.~Saito, J.~Nakamura, A.~Natori, Phys. Rev. B \textbf{76}, 115409 (2007)

\bibitem{jwang07}
J.~Wang, J.S. Wang, Appl. Phys. Lett. \textbf{90}, 241908 (2007)

\bibitem{jwangjp07}
J.~Wang, J.S. Wang, J. Phys.:Condens. Matter \textbf{19}, 236211 (2007)

\bibitem{wzhangjht07}
W.~Zhang, T.S. Fisher, N.~Mingo, J. Heat Transf. \textbf{129}, 483 (2007)

\bibitem{wzhangprb07}
W.~Zhang, N.~Mingo, T.S. Fisher, Phys. Rev. B \textbf{76}, 195429 (2007)

\bibitem{morooka08}
M.~Morooka, T.~Yamamoto, K.~Watanabe, Phys. Rev. B \textbf{77}, 033412 (2008)

\bibitem{mingo08}
N.~Mingo, D.A. Stewart, D.A. Broido, D.~Srivastava, Phys. Rev. B \textbf{77},
  033418 (2008)

\bibitem{patton01}
K.R. Patton, M.R. Geller, Phys. Rev. B \textbf{64}, 155320 (2001)

\bibitem{jwangapl06}
J.~Wang, J.S. Wang, Appl. Phys. Lett. \textbf{88}, 111909 (2006)

\bibitem{dattabook1}
S.~Datta, \emph{Electronic Transport in Mesoscopic Systems} (Cambridge Univ.
  Press, 1995)

\bibitem{qrzheng02}
Q.~Zheng, G.~Su, J.~Wang, H.~Guo, Eur. Phys. J. B \textbf{25}, 233 (2002)

\bibitem{callaway59}
J.~Callaway, Phys. Rev. \textbf{113}, 1046 (1959)

\bibitem{roufosse73}
M.~Roufosse, P.G. Klemens, Phys. Rev. B \textbf{7}, 5379 (1973)

\bibitem{ecsedy77}
D.J. Ecsedy, P.G. Klemens, Phys. Rev. B \textbf{15}, 5957 (1977)

\bibitem{yxiao03}
Y.~Xiao, X.H. Yan, J.X. Cao, J.W. Ding, J. Phys. Condens. Matter \textbf{15},
  L341 (2003)

\bibitem{jxcao04}
J.X. Cao, X.H. Yan, Y.~Xiao, J.W. Ding, Phys. Rev. B \textbf{69}, 073407 (2004)

\bibitem{hepplestone06}
S.P. Hepplestone, G.P. Srivastava, Phys. Rev. B \textbf{74}, 165420 (2006)

\bibitem{hepplestone07}
S.P. Hepplestone, G.P. Srivastava, J. Phys.: conf. ser. \textbf{61}, 414 (2007)

\bibitem{yunfenggu07}
Y.~Gu, Y.~Chen, Phys. Rev. B \textbf{76}, 134110 (2007)

\bibitem{schwinger61}
J.~Schwinger, J. Math. Phys. (New York) \textbf{2}, 407 (1961)

\bibitem{kadanoff62}
L..P. Kadanoff, G.~Baym, \emph{Quantum Statistical Mechanics}
  (Benjamin/Cummings, 1962)

\bibitem{keldysh65}
L.V. Keldysh, Soviet Phys. JETP \textbf{20}, 1018 (1965)

\bibitem{bonitz00}
M.~Bonitz, ed., \emph{Progress in Nonequilibrium Green's Functions} (World
  Scientific, 2000)

\bibitem{bonitz03}
M.~Bonitz, D.~Semkat, eds., \emph{Progress in Nonequilibrium Green's Functions
  II} (World Scientific, 2003)

\bibitem{wagner91}
M.~Wagner, Phys. Rev. B \textbf{44}, 6104 (1991)

\bibitem{jauho94}
A.P. Jauho, N.S. Wingreen, Y.~Meir, Phys. Rev. B \textbf{50}, 5528 (1994)

\bibitem{haug96}
H.~Haug, A.P. Jauho, \emph{Quantum Kinetics in Transport and Optics of
  Semiconductors} (Springer, 1996)

\bibitem{ciraci01}
S.~Ciraci, A.~Buldum, I.P. Batra, J. Phys.:Condens. Matter \textbf{13}, R537
  (2001)

\bibitem{mingo06}
N.~Mingo, Phys. Rev. B \textbf{74}, 125402 (2006)

\bibitem{hpliu06}
H.P. Liu, L.~Yi, Chin. Phys. Lett. \textbf{23}, 3194 (2006)

\bibitem{doniach74}
S.~Doniach, E.H. Sondheimer, \emph{Green's Functions for Solid State
  Physicists} (W. A. Benjamin, 1974)

\bibitem{mahan00}
G.D. Mahan, \emph{Many-Particle Physics}, 3rd~edn. (Kluwer Academic, 2000)

\bibitem{bruus04}
H.~Bruus, K.~Flensberg, \emph{Many-Body Quantum Theory in Condensed Matter
  Physics, an introduction} (Oxford Univ. Press, 2004)

\bibitem{langreth76}
D.C. Langreth, in \emph{Linear and Nonlinear Electron Transport in Solids},
  edited by J.T. Devreese, E.~van Doren (Plenum, 1976), pp. 3--32

\bibitem{nzeng07}
N.~Zeng, Ph.D. thesis, National Univ. Singapore (2007)

\bibitem{abrikosov63}
A.A. Abrikosov, L.P. Gorkov, I.E. Dzyaloshinski, \emph{Methods of Quantum Field
  Theory in Statistical Physics} (Dover Publ., 1963)

\bibitem{fetter71}
A.L. Fetter, J.D. Walecka, \emph{Quantum Theory of Many-Particle Systems}
  (McGraw-Hill, 1971)

\bibitem{zagoskin98}
A.M. Zagoskin, \emph{Quantum Theory of Many-Body Systems} (Springer, 1998)

\bibitem{amit84}
D.J. Amit, \emph{Field Theory, the Renormalization Group, and Critical
  Phenomena}, 2nd~edn. (World Scientific, 1984)

\bibitem{procacci92}
P.~Procacci, G.~Cardini, R.~Righini, S.~Califano, Phys. Rev. B \textbf{45},
  2113 (1992)

\bibitem{valle92}
R.G.D. Valle, P.~Procacci, Phys. Rev. B \textbf{46}, 6141 (1992)

\bibitem{valle00}
R.G.D. Valle, P.~Procacci, J. Comput. Phys. \textbf{165}, 428 (2000)

\bibitem{niu99}
C.~Niu, D.L. Lin, T.H. Lin, J. Phys.:Condens. Matter \textbf{11}, 1511 (1999)

\bibitem{leeuwen06}
R.~van Leeuwen, N.E. Dahlen, G.~Stefanucci, C.O. Almbladh, U.~von Barth, in
  \emph{Time-Dependent Density Functional Theory}, edited by M.A.L. Marques,
  et~al. (Springer, 2006), chap.~3

\bibitem{thygesen07}
K.S. Thygesen, A.~Rubio, J. Chem. Phys. \textbf{126}, 091101 (2007)

\bibitem{thygesen-0710.0482}
K.S. Thygesen, A.~Rubio, arXiv:0710.0482

\bibitem{baym62}
G.~Baym, Phys. Rev. \textbf{127}, 1391 (1962)

\bibitem{czwang90}
C.Z. Wang, C.T. Chan, K.M. Ho, Phys. Rev. B \textbf{42}, 11276 (1990)

\bibitem{yhlee91}
Y.H. Lee, R.~Biswas, C.M. Soukoulis, C.Z. Wang, C.T. Chan, K.M. Ho, Phys. Rev.
  B \textbf{43}, 6573 (1991)

\bibitem{juli98}
J.~Li, L.~Porter, S.~Yip, J. Nucl. Mat. \textbf{255}, 139 (1998)

\bibitem{jscao93}
J.~Cao, G.A. Voth, J. Chem. Phys. \textbf{99}, 10070 (1993)

\bibitem{yonetani04}
Y.~Yonetani, K.~Kinugawa, J. Chem. Phys. \textbf{120}, 10624 (2004)

\bibitem{jswang07}
J.S. Wang, Phys. Rev. Lett. \textbf{99}, 160601 (2007)

\bibitem{lindenberg90}
K.~Lindenberg, B.J. West, \emph{The Nonequilibrium Statistical Mechanics of
  Open and Closed Systems} (VCH, 1990)

\bibitem{haanggi05}
P.~H{\"a}nggi, G.L. Ingold, Chaos \textbf{15}, 026105 (2005)

\bibitem{dharprb06}
A.~Dhar, D.~Sen, Phys. Rev. B \textbf{73}, 085119 (2006)

\bibitem{schmid82}
A.~Schmid, J. Low Temp. Phys. \textbf{49}, 609 (1982)

\bibitem{weiss99}
U.~Weiss, \emph{Quantum Dissipative Systems}, 2nd~edn. (World Scientific, 1999)

\bibitem{jtlu08}
J.T. L\"{u}, J.S. Wang, arXiv:08xx.xxxx

\bibitem{balletine98}
L.E. Ballentine, S.M. McRae, Phys. Rev. A \textbf{58}, 1799 (1998)

\bibitem{prezhdo00}
O.V. Prezhdo, Y.V. Pereverzev, J. Chem. Phys. \textbf{113}, 6557 (2000)

\bibitem{pahl02}
E.~Pahl, O.V. Prezhdo, J. Chem. Phys. \textbf{116}, 8704 (2002)

\bibitem{heatwole05}
E.M. Heatwole, O.V. Prezhdo, J. Chem. Phys. \textbf{122}, 234109 (2005)

\bibitem{prezhdo06}
O.V. Prezhdo, Theor. Chem. Acc. \textbf{116}, 206 (2006)

\bibitem{barik03}
D.~Barik, B.C. Bag, D.S. Ray, J. Chem. Phys. \textbf{119}, 12973 (2003)

\bibitem{barik04}
D.~Barik, D.S. Ray, J. Chem. Phys. \textbf{121}, 1681 (2004)

\bibitem{horsfield04b}
A.P. Horsfield, D.R. Bowler, A.J. Fisher, T.N. Todorov, C.G. S\'anchez, J.
  Phys.:Condens. Matter \textbf{16}, 8251 (2004)

\bibitem{landau05}
D.P. Landau, K.~Binder, \emph{A Guide to Monte Carlo Simulations in Statistical
  Physics}, 2nd~edn. (Cambridge Univ. Press, 2005)

\bibitem{michel05}
M.~Michel, G.~Mahler, J.~Gemmer, Phys. Rev. Lett. \textbf{95}, 180602 (2005)

\bibitem{michel06}
M.~Michel, J.~Gemmer, G.~Mahler, Phys. Rev. E \textbf{73}, 016101 (2006)

\bibitem{michelrev06}
M.~Michel, J.~Gemmer, G.~Mahler, Int. J. Mod. Phys. B \textbf{20}, 4855 (2006)

\bibitem{segal05}
D.~Segal, A.~Nitzan, Phys. Rev. Lett. \textbf{94}, 034301 (2005)

\bibitem{segaljcp05}
D.~Segal, A.~Nitzan, J. Chem. Phys. \textbf{122}, 194704 (2005)

\bibitem{segal06}
D.~Segal, Phys. Rev. B \textbf{73}, 205415 (2006)

\bibitem{segal-arxiv07}
L.A. Wu, D.~Segal, arXiv:0711.4599

\bibitem{steinigeweg07}
R.~Steinigeweg, H.P. Breuer, J.~Gemmer, Phys. Rev. Lett. \textbf{99}, 150601
  (2007)

\bibitem{buldum99}
A.~Buldum, D.M. Leitner, S.~Ciraci, Europhys. Lett. \textbf{47}, 208 (1999)

\bibitem{leitner00}
D.M. Leitner, P.G. Wolynes, Phys. Rev. E \textbf{61}, 2902 (2000)

\bibitem{leitner01}
D.M. Leitner, Phys. Rev. B \textbf{64}, 094201 (2001)

\bibitem{santhosh07}
G.~Santhosh, D.~Kumar, Phys. Rev. E \textbf{76}, 021105 (2007)

\bibitem{mitra04}
A.~Mitra, I.~Aleiner, A.J. Millis, Phys. Rev. B \textbf{69}, 245302 (2004)

\bibitem{horsfield04}
A.P. Horsfield, D.R. Bowler, A.J. Fisher, T.N. Todorov, M.J. Montgomery, J.
  Phys.: Condens. Matter \textbf{16}, 3609 (2004)

\bibitem{todorov98}
T.N. Todorov, Phil. Mag. B \textbf{77}, 965 (1998)

\bibitem{agrait}
N.~Agra\"it, C.~Untiedt, G.~Rubio-Bollinger, S.~Vieira, Phys. Rev. Lett.
  \textbf{88}, 216803 (2002)

\bibitem{segal:3915}
D.~Segal, A.~Nitzan, J. Chem. Phys. \textbf{117}, 3915 (2002)

\bibitem{montgomery02}
M.J. Montgomery, T.N. Todorov, A.P. Sutton, J. Phys.: Condens. Matter
  \textbf{14}, 5377 (2002)

\bibitem{horsfield05}
A.P. Horsfield, D.R. Bowler, A.J. Fisher, T.N. Todorov, C.G. S\'{a}nchez, J.
  Phys.: Condens. Matter \textbf{17}, 4793 (2005)

\bibitem{horsfield06}
A.P. Horsfield, D.R. Bowler, H.~Ness, C.G. S\'anchez, T.N. Todorov, A.J.
  Fisher, Rep. Prog. Phys. \textbf{69}, 1195 (2006)

\bibitem{ycchen03}
Y.C. Chen, M.~Zwolak, M.D. Ventra, Nano Letters \textbf{3}, 1691 (2003)

\bibitem{frederiksen04}
T.~Frederiksen, M.~Brandbyge, N.~Lorente, A.P. Jauho, Phys. Rev. Lett.
  \textbf{93}, 256601 (2004)

\bibitem{pecchia07}
A.~Pecchia, G.~Romano, A.D. Carlo, Phys. Rev. B \textbf{75}, 035401 (2007)

\bibitem{sun07}
Q.F. Sun, X.C. Xie, Phys. Rev. B \textbf{75}, 155306 (2007)

\bibitem{lu07}
J.T. L\"{u}, J.S. Wang, Phys. Rev. B \textbf{76}, 165418 (2007)

\bibitem{tighe97}
T.S. Tighe, J.M. Worlock, M.L. Roukes, Appl. Phys. Lett. \textbf{70}, 2687
  (1997)

\bibitem{roukes99}
M.L. Roukes, Physica B \textbf{263}, 1 (1999)

\bibitem{hone99}
J.~Hone, M.~Whitney, C.~Piskoti, A.~Zettl, Phys. Rev. B \textbf{59}, R2514
  (1999)

\bibitem{djyang02}
D.J. Yang, Q.~Zhang, G.~Chen, S.F. Yoon, J.~Ahn, S.G. Wang, Q.~Zhou, Q.~Wang,
  J.Q. Li, Phys. Rev. B \textbf{66}, 165440 (2002)

\bibitem{hqxie07}
H.~Xie, A.~Cai, X.~Wang, Phys. Lett. A \textbf{369}, 120 (2007)

\bibitem{berber00}
S.~Berber, Y.K. Kwon, D.~Tom\'anek, Phys. Rev. Lett. \textbf{84}, 4613 (2000)

\bibitem{lukes07}
J.R. Lukes, H.~Zhong, J. Heat Transf. \textbf{129}, 705 (2007)

\bibitem{pkim02}
P.~Kim, L.~Shi, A.~Majumdar, P.L. McEuen, Phys. Rev. Lett. \textbf{87}, 215502
  (2001)

\bibitem{pkim02b}
P.~Kim, L.~Shi, A.~Majumdar, P.L. McEuen, Physica B \textbf{323}, 67 (2002)

\bibitem{cyu05}
C.~Yu, L.~Shi, Z.~Yao, D.~Li, A.~Majumdar, Nano Lett. \textbf{5}, 1842 (2005)

\bibitem{pop06}
E.~Pop, D.~Mann, Q.~Wang, K.~Goodson, H.~Dai, Nano Lett. \textbf{6}, 96 (2006)

\bibitem{zlwang07}
Z.L. Wang, D.W. Tang, X.B. Li, X.H. Zheng, W.G. Zhang, L.X. Zheng, Y.T. Zhu,
  A.Z. Jin, H.F. Yang, C.Z. Gu, Appl. Phys. Lett. \textbf{91}, 123119 (2007)

\bibitem{hychiu05}
H.Y. Chiu, V.V. Deshpande, H.W.C. Postma, C.N. Lau, C.~Mik\'o, L.~Forr\'o,
  M.~Bockrath, Phys. Rev. Lett. \textbf{95}, 226101 (2005)

\bibitem{dyli03}
D.~Li, Y.~Wu, P.~Kim, L.~Shi, P.~Yang, A.~Majumdar, Appl. Phys. Lett.
  \textbf{83}, 2934 (2003)

\bibitem{bourgeois07}
O.~Bourgeois, T.~Fournier, J.~Chaussy, J. Appl. Phys. \textbf{101}, 016104
  (2007)

\bibitem{hochbaum08}
A.I. Hochbaum, R.~Chen, R.D. Delgado, W.~Liang, E.C. Garnett, M.~Najarian,
  A.~Majumdar, P.~Yang, Nature \textbf{451}, 163 (2008)

\bibitem{cwchang06}
C.W. Chang, D.~Okawa, A.~Majumdar, A.~Zettl, Science \textbf{314}, 1121 (2006)

\bibitem{cwchang07}
C.W. Chang, D.~Okawa, H.~Garcia, T.D. Yuzvinsky, A.~Majumdar, A.~Zettl, Appl.
  Phys. Lett. \textbf{90}, 193114 (2007)

\bibitem{cwwang07prl}
C.W. Chang, D.~Okawa, H.~Garcia, A.~Majumdar, A.~Zettl, Phys. Rev. Lett.
  \textbf{99}, 045901 (2007)

\bibitem{feher07}
A.~Feher, S.A. Egupov, A.G. Shkorbatov, Low Temp. Phys. \textbf{33}, 861 (2007)

\bibitem{zbge06}
Z.~Ge, D.G. Cahill, P.V. Braun, Phys. Rev. Lett. \textbf{96}, 186101 (2006)

\bibitem{nitzan07}
A.~Nitzan, Science \textbf{317}, 759 (2007)

\bibitem{zhwang07}
Z.~Wang, J.A. Carter, A.~Lagutchev, Y.K. Koh, N.H. Seong, D.G. Cahill, D.D.
  Dlott, Science \textbf{317}, 787 (2007)

\bibitem{negfweb}
Http://staff.science.nus.edu.sg/\char`~phywjs/NEGF/negf.html

\end{thebibliography}
%
%
%

\end{document}